\documentclass[preprint,12pt]{elsarticle}

\usepackage{amssymb}
\usepackage{epsfig}
\usepackage{graphics}
\usepackage{graphicx}
\usepackage{subfigure}
\usepackage{exscale}
\usepackage{latexsym}
\usepackage{makeidx}
\usepackage[section]{placeins}
\usepackage{xtab}
\usepackage{epsf}
\usepackage{bbm}

\usepackage{lineno}

\journal{Nuclear Instruments and Methods in Physics Research Section A}

\begin{document}
\begin{frontmatter}

\title{Characterization of LAPPD timing at CERN PS testbeam}

\author[INFN-TS]{Deb Sankar Bhattacharya}
\author[INFN-TS, UNI-TS]{Andrea Bressan}
\author[INFN-TS]{Chandradoy Chatterjee}
\author[INFN-TS]{Silvia Dalla Torre}
\author[INFN-TS]{Mauro Gregori}
\author[BNL]{Alexander Kiselev}
\author[INFN-TS]{Stefano Levorato}
\author[INFN-TS, UNI-TS]{Anna Martin}
\author[INFN-GE]{Saverio Minutoli}
\author[INFN-GE]{Mikhail Osipenko\corref{cor1}}
\ead{osipenko@ge.infn.it}
\author[INFN-TS]{Richa Rai}
\author[INFN-GE]{Marco Ripani}
\author[INFN-TS]{Fulvio Tessarotto}
\author[INFN-TS]{Triloki Triloki}

\cortext[cor1]{Please address correspondence to Mikhail Osipenko}

\address[INFN-TS]{INFN Sezione di Trieste, Trieste, 34127 Italy}
\address[UNI-TS]{INFN Universit\`a di Trieste, Trieste, 34127 Italy}
\address[BNL]{Brookhaven National Lab, Upton, NY 11973, USA}
\address[INFN-GE]{INFN, Sezione di Genova, Genova, 16146 Italy}

\begin{abstract}
Large Area Picosecond PhotoDetectors (LAPPDs) are photosensors based 
on microchannel plate technology with about 400~cm$^2$ sensitive area. The external readout plane of a capacitively coupled LAPPD can be segmented into pads providing a 
spatial resolution down to 1~mm scale. The LAPPD signals have about 0.5~ns risetime followed by a slightly longer falltime
and their amplitude reaches a few dozens of mV per single photoelectron.
\par
In this article, we report on the measurement of the time resolution 
of an LAPPD prototype in a test beam exercise at CERN PS. Most of the previous measurements of LAPPD time resolution had been performed with laser sources. In this article we report  time resolution  measurements obtained through the detection of Cherenkov radiation emitted by high energy hadrons.
Our approach has been demonstrated capable of measuring time resolutions as fine as 25-30~ps.
The available prototype had performance limitations, which prevented us from applying the 
optimal high voltage setting. The measured time resolution 
for single photoelectrons is about 80~ps r.m.s.

\end{abstract}


\end{frontmatter}

\section{Introduction}\label{sec:intro}
Low noise photodetectors with single photoelectron detection capability, high Quantum Efficiency (QE) 
and long lifetime  are needed for fundamental research in particle and nuclear physics. In particular, they are 
required for Cherenkov imaging devices, including Ring Imaging CHerenkov detectors (RICH) and Detection of Internally Reflected Cherenkov light detectors (DIRC). Conventional PhotoMultiplier Tubes (PMT) are vacuum-based sensors matching these characteristics. 
However, they cannot be operated in presence of a strong magnetic field without a bulky shielding, and their single photon timing resolution is limited. 
\par
MicroChannel Plates-PMT (MCP-PMT) are vacuum-based photomultipliers introduced to overcome the PMT limitations. They are based on the concept of a continuous dynode for 
electron multiplication, a concept introduced almost one century ago~\cite{Farnsworth}. Two other essential ingredients must be mentioned: (i) the secondary electron emission, a field developed for the PMTs, detectors introduced in the late 30's of the twentieth century, and (ii) the grouping of very small, tubular continuous dynode channels (pores) in parallel arrays, first considered in the 1960's ~\cite{Goodrich}. Nowadays the small tubular approach has evolved in plates of capillaries. The MCP-based detector concept was introduced in 1979~\cite{Wiza}. A long evolution followed to overcome ageing effects, mostly related to 
ion bombardment of the photocathode, and improvement of the overall MCP-PMT performance. At present, MCP-PMTs are characterized by high gain of O(10$^6$) and more, very fine time resolution, well below 100~ps for single photoelectrons, good radiation hardness, showing no or limited ageing up to 100-1000~mC of extracted charge per cm$^2$ and more, good capability of operation in magnetic field of 1~T and more, in particular when small pore diameter is selected (O(10~$\mu$m)). They are nowadays commercially available from Hamamatsu\footnote{Hamamatsu Photonis, 325-6, Sunayama-cho, Naka-ku, Hamamatsu City, Shizuoka Pref., 430-8587, Japan}, Photek\footnote{Photek, 26 Castleham Road, St Leonards on SeaEast Sussex, TN38 9NS, United Kingdom} and Photonis\footnote{Photonis Group S.A.S., 18 Avenue Pythagore, 33700 Mérignac, France}.  They can be both single channel and multichannel devices with active surface up to a few square inches. It is relevant to underline that the increase in the detector lifetime is largely due to the application of ultra-thin films of resistive and emissive layers by Atomic Layer Deposition (ALD) technique, initially  developed by Arradiance Inc.\footnote{Arradiance Inc., 142 North Road, Sudbury, MA 01776, USA} \cite{Beaulieu_1,Beaulieu_2}.
\par
The very fine time resolution that MCP-PMTs offer, makes them of interest also for applications beyond Cherenkov imaging detectors, such as Time-Of-Flight (TOF) measurements, timing layers in calorimetry and medical Positron Emission Tomography (PET).
\par
Large Area Picosecond Photo-Detectors (LAPPD) are large area (O(20$\times$20~cm$^2$)) MCP-based detectors~\cite{lappd_intro,advance_lappd,shin_lappd},
resulted from a combined effort of academia and industry. LAPPDs 
are presently produced by Incom Inc\footnote{Incom Inc., 294 Southbridge Rd, Charlton, Massachusetts, 01507, United States}.
Thanks to the large active surface, LAPPDs offer the advantage of a large active area fraction and reduced cost per surface unit. They are already used in the ANNIE neutrino experiment~\cite{annie} and considered for a number of future projects. The present assessment of the LAPPD R\&D, characterization studies and application options~\cite{lappd_time_p,lubliana_lappd,cherenkov_lappd} has been reviewed in three recent workshops \cite{LAPPD_Workshop_1, LAPPD_Workshop_2, LAPPD_Workshop_3}.
\par
Our interest is related to the proposed use of LAPPDs in the ePIC experiment at the US Electron Ion Collider (EIC)~\cite{white_paper,yellow_report}, where 
LAPPDs are considered 
for the proximity focusing RICH in the backward endcap as well as the barrel DIRC. In the backward endcap RICH, the LAPPDs will 
be placed in the detector acceptance. Therefore, they can also provide TOF information detecting the Cherenkov light produced in the sensor
window by the through-going particles. 
\par
In this article, we report on the measurement of the time resolution of an LAPPD prototype obtained from  a 
test beam at CERN PS. Most of the previous measurements of LAPPD time resolution had been performed with laser sources.
In this article we report time resolution measurements obtained by detecting Cherenkov radiation emitted by high energy hadrons.


\section{The LAPPD sensor}\label{sec:LAPPD_unit}
\begin{figure}[!ht]
\begin{center}
\includegraphics[width=0.95\textwidth]{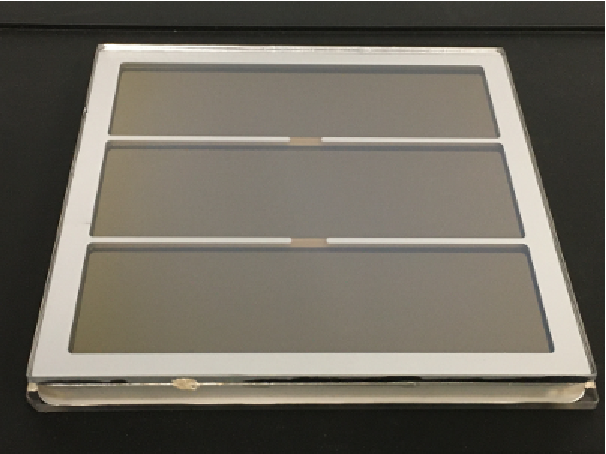}
\caption{\label{fig:LAPPD_photo} A photograph of the LAPPD 124 [picture courtesy: Incom]. }
\end{center}
\end{figure}

The studies reported in this article have been performed using the LAPPD unit no.~124 by Incom. It has an active surface of 192$\times$192~mm$^2$, partially 
obscured by a dead area due to the
spacers
(Fig.~\ref{fig:LAPPD_photo}).

\par
This LAPPD is a unit of type "Generation II", namely it is characterized by a resistive anode realized by coating the 
rear plate closing the sensor 
vacuum volume with a thin Cr layer. 
The capacitively coupled readout plane is designed on the readout PCB in a form of an array of 8$\times$8 square pads of 1~inch size. The connections from the readout pads were adapted to 50~$\Omega$ impedance and transformed into the differential ones and back through a pair of RF-transformers for a better suppression of the environmental noise.
\par
\begin{figure}[!ht]
\begin{center}
\includegraphics[width=0.95\textwidth]{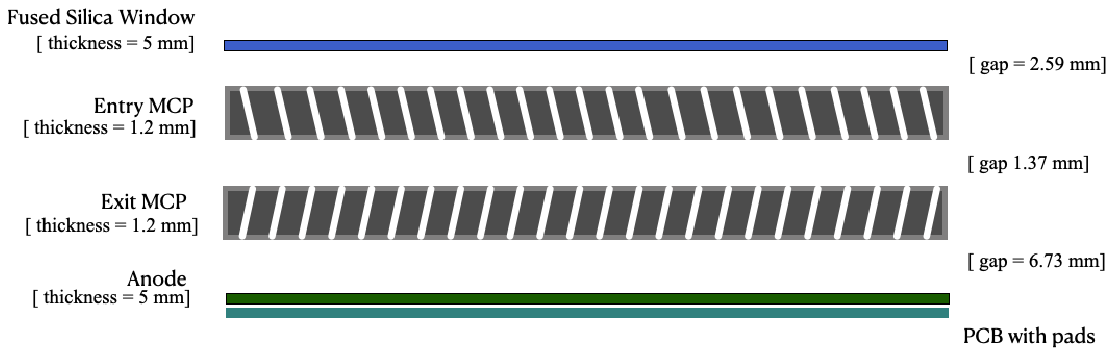}
\caption{\label{fig:LAPPD} Schematics of the LAPPD no.~124 architecture (not to scale). }
\end{center}
\end{figure}

The main components of the sensor, other than the anode, are the bialkali PhotoCathode (PC), 
deposited on the inner 
side of the fused silica entrance window, and two layers of MCPs with 20~$\mu$m-diameter pores.
Figure~\ref{fig:LAPPD} presents a schematics of the LAPPD architecture providing information about the key geometrical parameters. 
\par
The LAPPD no.~124 used in our studies was affected by a
substantial spontaneous electron emission from the photocathode, in particular observed after sensor maintenance 
which 
required sensor 
exposure to ambient illumination. Stable operating conditions could be obtained only after a long conditioning process (O(1d)) keeping the sensor in the dark box and applying progressively increasing bias voltages to the MCPs. This feature also posed a limitation to the maximum gain at which the LAPPD could be operated.
\par
The LAPPD charge multiplication gain and collection efficiency were determined by the voltage differences 
between its five electrodes. Although, the two surfaces of each MCP layer were not independent electrodes: they were connected by the intrinsic resistivity of the MCP capillaries. Therefore, no steadily defined electrical bias could be obtained using High Voltage (HV) power supply units with common grounding. These considerations have guided our choice of the power supply device.
\par
\begin{figure}[!ht]
\begin{center}
\includegraphics[width=0.95\textwidth]{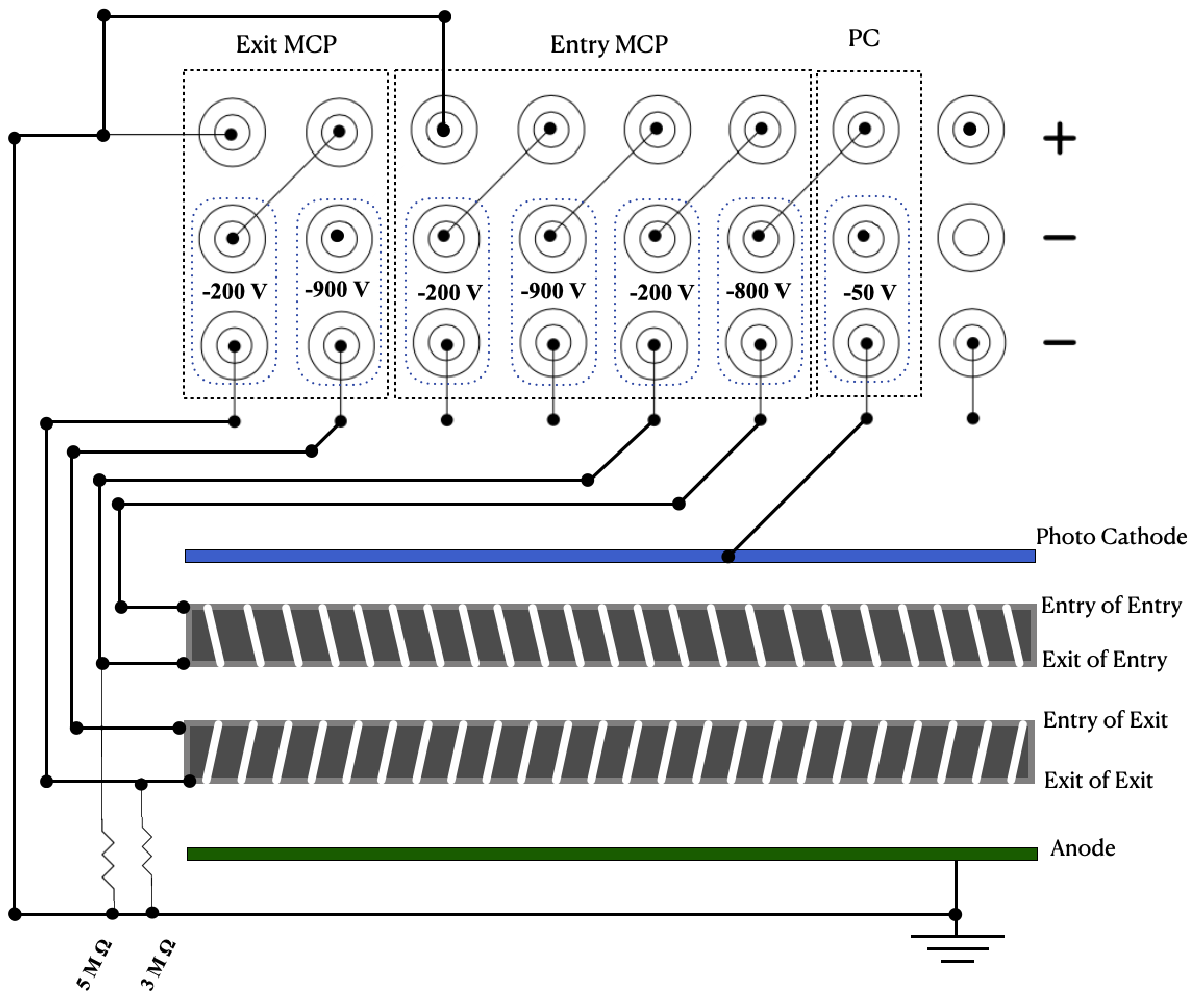}
\caption{\label{fig:HV_schematic} The scheme used to power the LAPPD: circles represent power supply SHV connectors. The  resistors connecting 
the exit electrodes of the two MCPs to ground are needed to guarantee that these electrodes are biased at the selected voltages.}
\end{center}
\end{figure}

A negative voltage bias configuration was used, with the Photo-Cathode (PC) at the maximum negative voltage and the anode at ground potential. The LAPPD HV bias voltages were supplied by the CAEN\footnote{CAEN S.p.A., Via della Vetraia, 11, 55049 Viareggio LU, Italy} 
DT1415ET power supply. This unit allowed for a HV channel stacking connection scheme (a daisy chain). 
Maximum bias voltages applicable to the LAPPD under test were limited by the current and the readout voltage instabilities, observed on the power supply on-line monitors. Figure~\ref{fig:HV_schematic} shows the HV connection scheme with the main voltage configuration: 50/800/200/900/200, where the numbers indicate differential HV values in volts in the following order: between the Photo-Cathode and the entry surface of MCP1, the first multiplication layer of the LAPPD across MCP1,  the transfer field guiding the electrons from MCP1 to MCP2, the second multiplication layer in MCP2 and the transfer field from MCP2 to the anode. The schematic also shows the absolute values of the voltages on the respective electrodes. The total voltage difference between the cathode and the anode was 2150~V, which is just 60~V below the maximum suggested by Incom. In order to give an idea about the underlying electric fields 
across the LAPPD vacuum gaps we report the distances: PC-MCP1 2.6~mm, MCP1-MCP2 1.4~mm and MCP2-anode 6.7~mm. Thus, the electric fields in the vacuum gaps are 0.19, 1.5 and 0.3 kV/cm, respectively. A large fraction of the data, including those runs for which we report the main results, have been collected removing the external 5~M$\Omega$ resistor, via which the "Exit of Entry" electrode is connected to ground. In this configuration the currents read out by the power supply monitor were about 9 nA in the PC-MCP1 gap, 150~$\mu$A across the MCP1, 0.7~$\mu$A in the MCP1-MCP2 gap, 233~$\mu$A across the MCP2 and 66~$\mu$A in the MCP2-anode gap (dominated by the 3~M$\Omega$ resistor).
%
%

%
%

%

\section{Measurement principle and Monte Carlo simulations}\label{sec:mc_sim}
The LAPPD timing properties 
in this work are studied by detecting Cherenkov light generated by the hadron beam in a plano-convex aspheric fused silica lens, acting as a radiator. The produced Cherenkov light, after a total internal reflection from the planar side of the lens, leaves the radiator and reaches the window of the LAPPD. The hadron beam also produces another Cherenkov 
flash when passing through the 5~mm thick LAPPD quartz window. Two configurations can be envisaged: (i) the radiator sitting upstream of the LAPPD positioned with its entrance window looking upstream or (ii) the radiator located downstream of the the sensor with its entrance window looking downstream. Both configurations have been studied and optimized in terms of the expected timing resolution using Geant4 Monte Carlo simulations. The most favorable one, namely (ii), has been selected. The simulation studies reported in the following refer to the selected layout. An acrylic plate used as an 
ultraviolet
(UV) light filter is interposed between the lens radiator and the sensor to reduce the overall number of Cherenkov photons from the radiator reaching the LAPPD photocathode and to reduce the Chromatic effects.
\par
\begin{figure}[!ht]
\begin{center}
\includegraphics[scale=0.25]{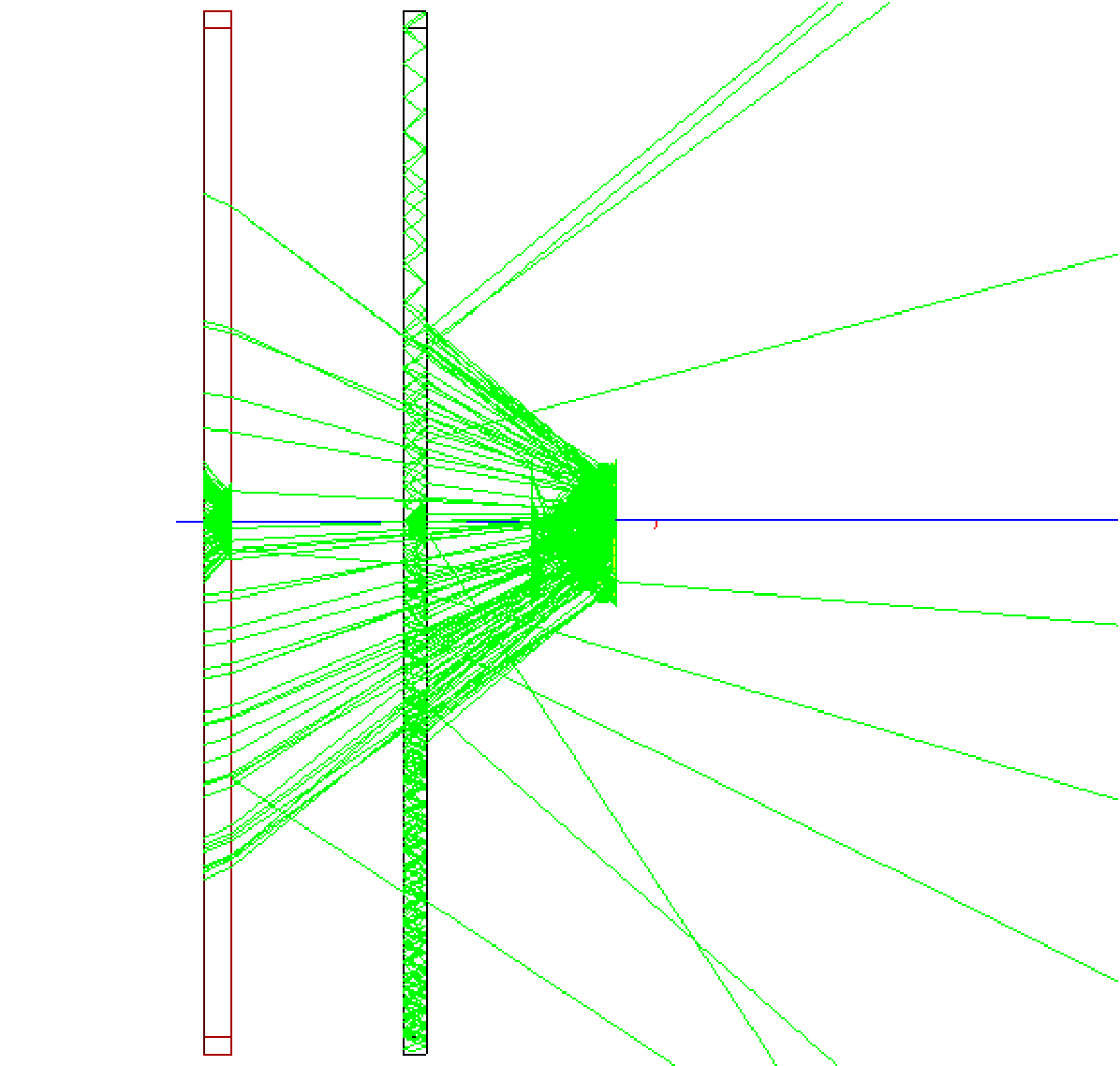}
\caption{\label{fig:geant4_view}Visualization of a Geant4 simulation: green lines represent optical photons, a red rectangle is the LAPPD fused silica window, a black rectangle is the Acrylic Filter and the yellow structure is the aspheric lens. The proton track is shown by a blue line and the direction of incoming particles is from left to right.
}
\end{center}
\end{figure}

The simulations included a description of a LAPPD fused silica window, a fused silica lens radiator and an acrylic filter as shown in Fig.~\ref{fig:geant4_view}. The Cherenkov radiation process and ray tracing were used to describe the light emission and propagation to the LAPPD photocathode. We used LAPPD Quantum Efficiency (QE) provided in the sensor data sheets to simulate the photo-electron emission, assuming a 100\% charge collection efficiency inside the LAPPD. The primary beam was simulated using 6~GeV/c protons, which is the mean momentum of the range available at the test beam line.

\begin{figure}[!ht]
\begin{center}
\includegraphics[scale=0.25, angle=270]{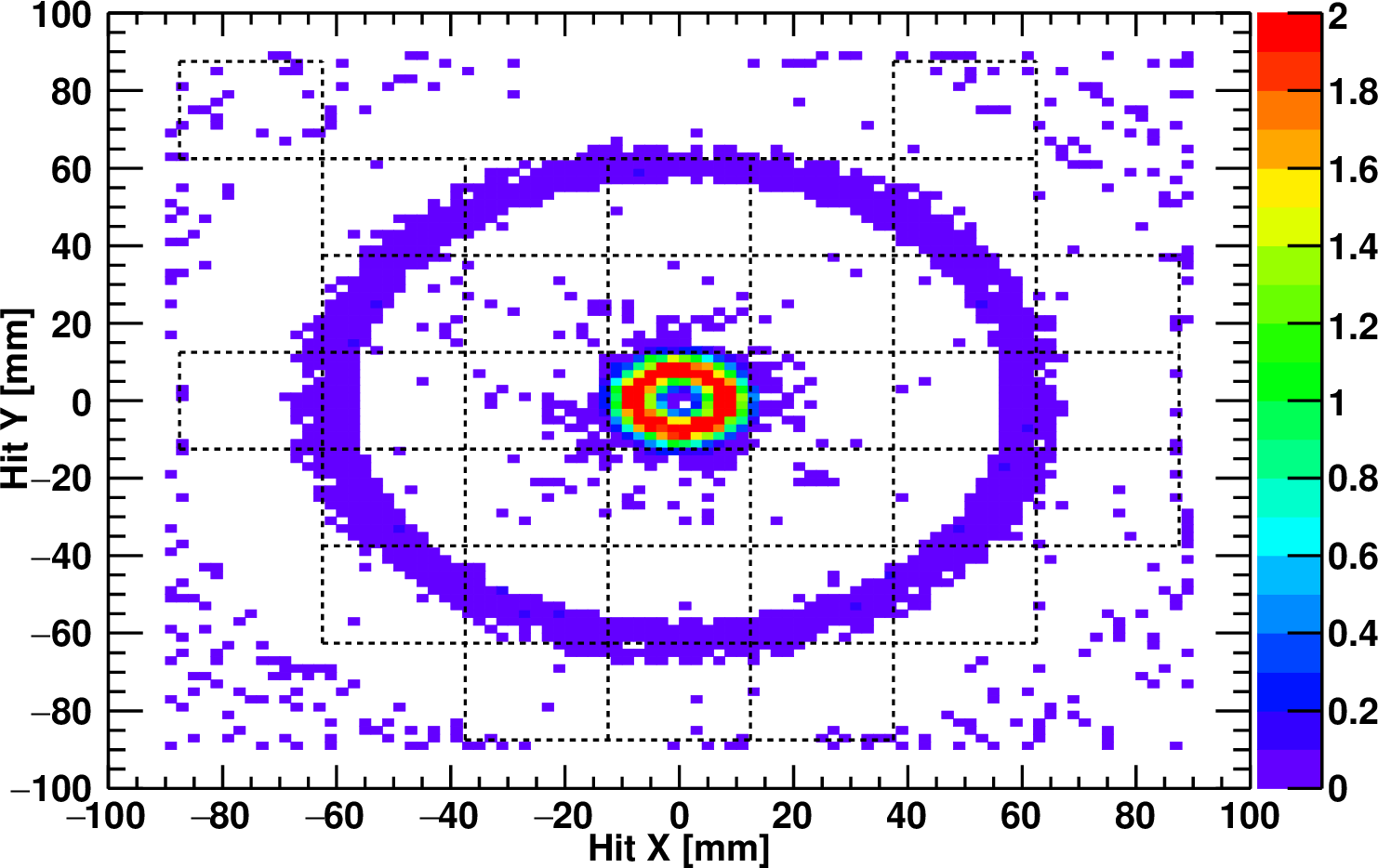}~%
\includegraphics[scale=0.25, angle=270]{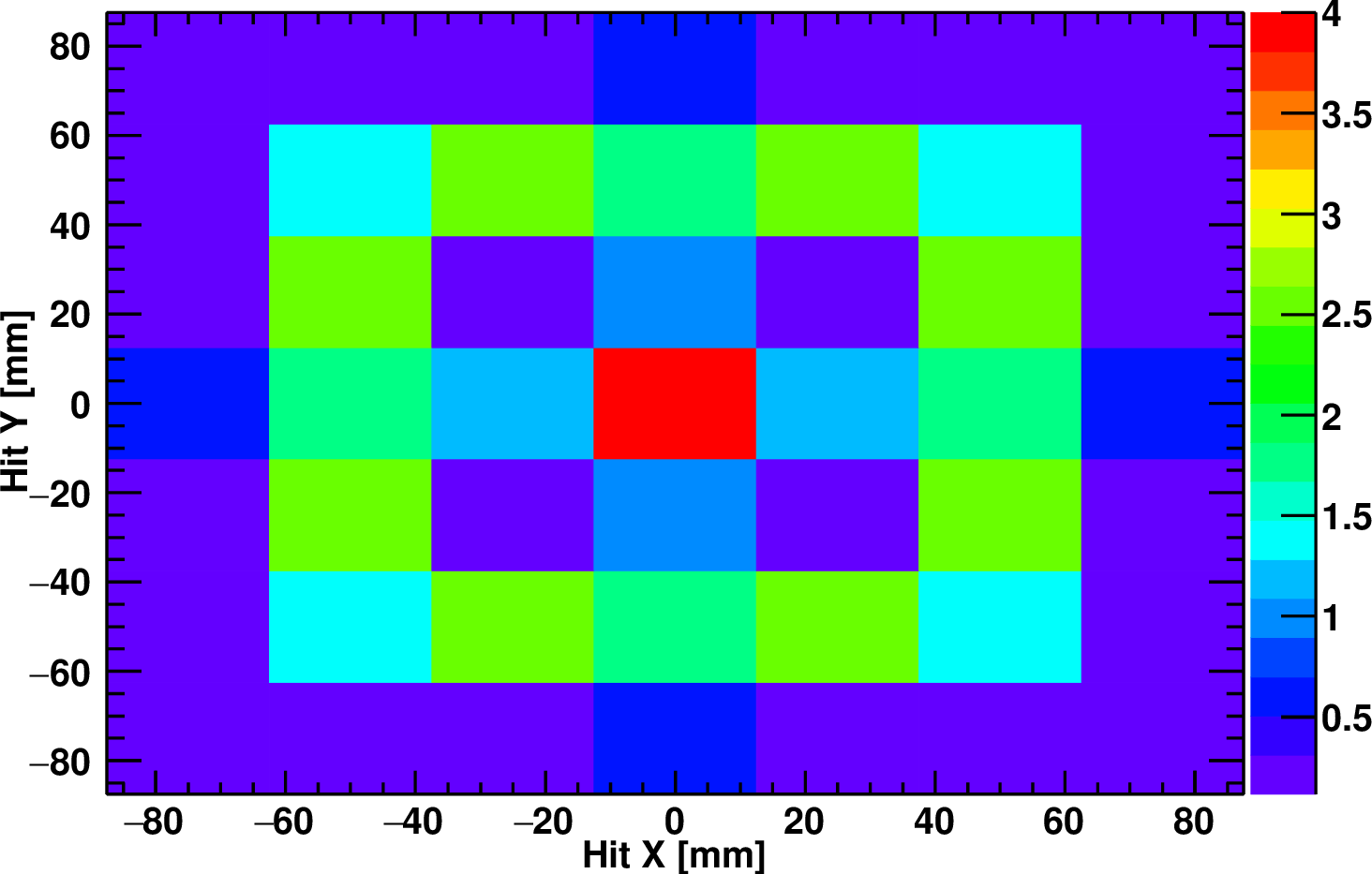}
\caption{\label{fig:geant4_npe_map}Mean number of photo-electrons produced from the LAPPD photocathode obtained from Geant4 simulations in 2~mm bins (left) and in 1 inch pads (right). In the left plot,  the borders of the pads used in our studies are shown by dashed lines.
The Z-axis maximum of 2 (left) and 4 (right) photo-electrons was applied in order to make the photons coming from the quartz lens and forming a Cherenkov ring image better visible.
}
\end{center}
\end{figure}

The 14~mm thick lens was placed downstream from LAPPD looking upstream with its convex side.
Good focusing of the Cherenkov ring photons was obtained for a distance of 60~mm between the LAPPD window and the center of the lens. The simulated number of photo-electrons per event is shown in Fig.~\ref{fig:geant4_npe_map}. The direct beam spot was overpopulated, reaching 179~PhotoElectrons (PE) in the central pad. Therefore, the maximum number of PEs per bin is limited in the plots of Fig.~\ref{fig:geant4_npe_map}
(see caption for more details).
In the experiment, we covered the window region corresponding to this pad with with a black tape coupled to the window by optical grease in order to reduce the number of detected PEs. The average number of photo-electrons in the pads collecting the photoelectrons from the Cherenkov ring was $\sim$2~PEs.

\begin{figure}[!ht]
\begin{center}
\includegraphics[scale=0.25, angle=270]{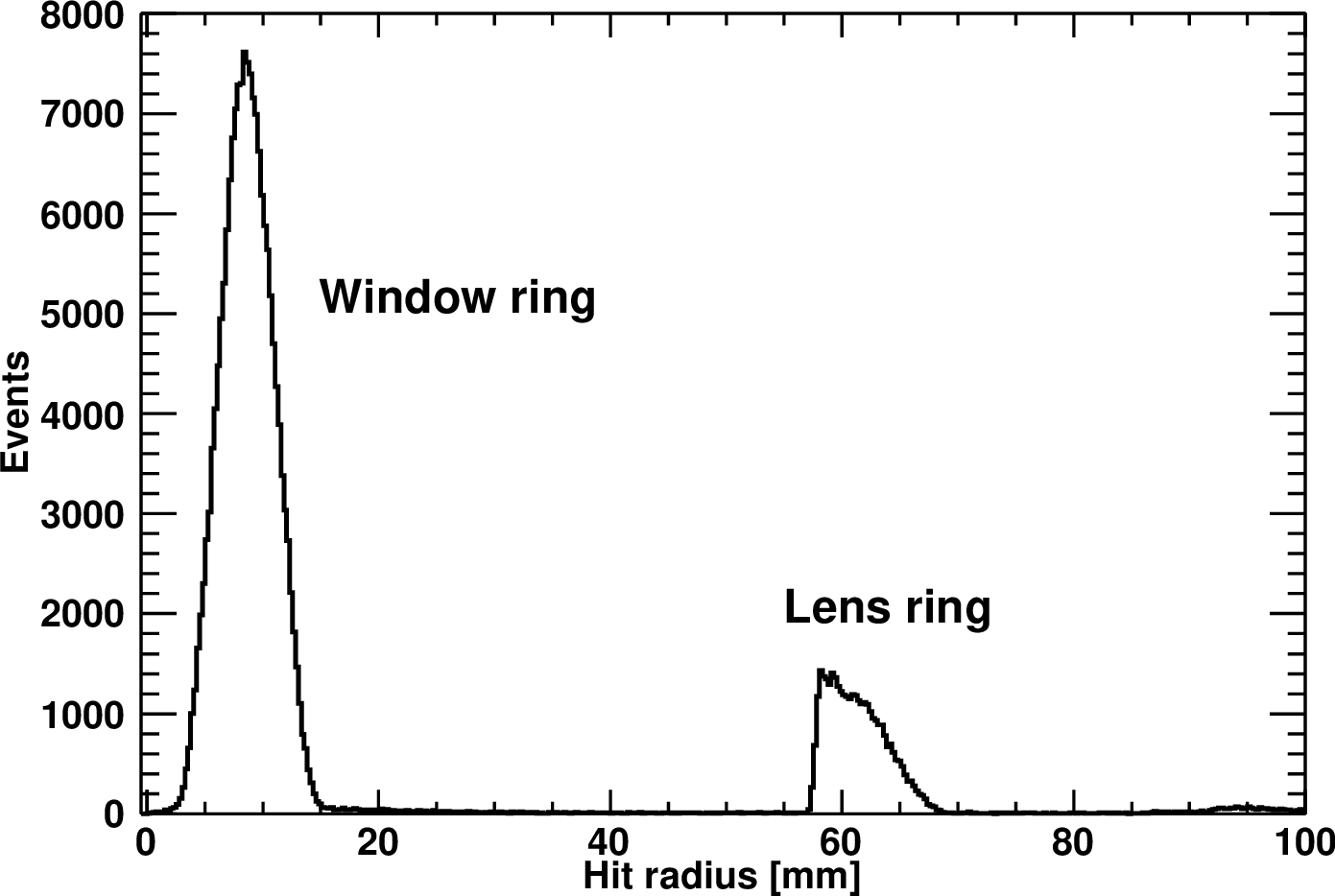}~%
\includegraphics[scale=0.25, angle=270]{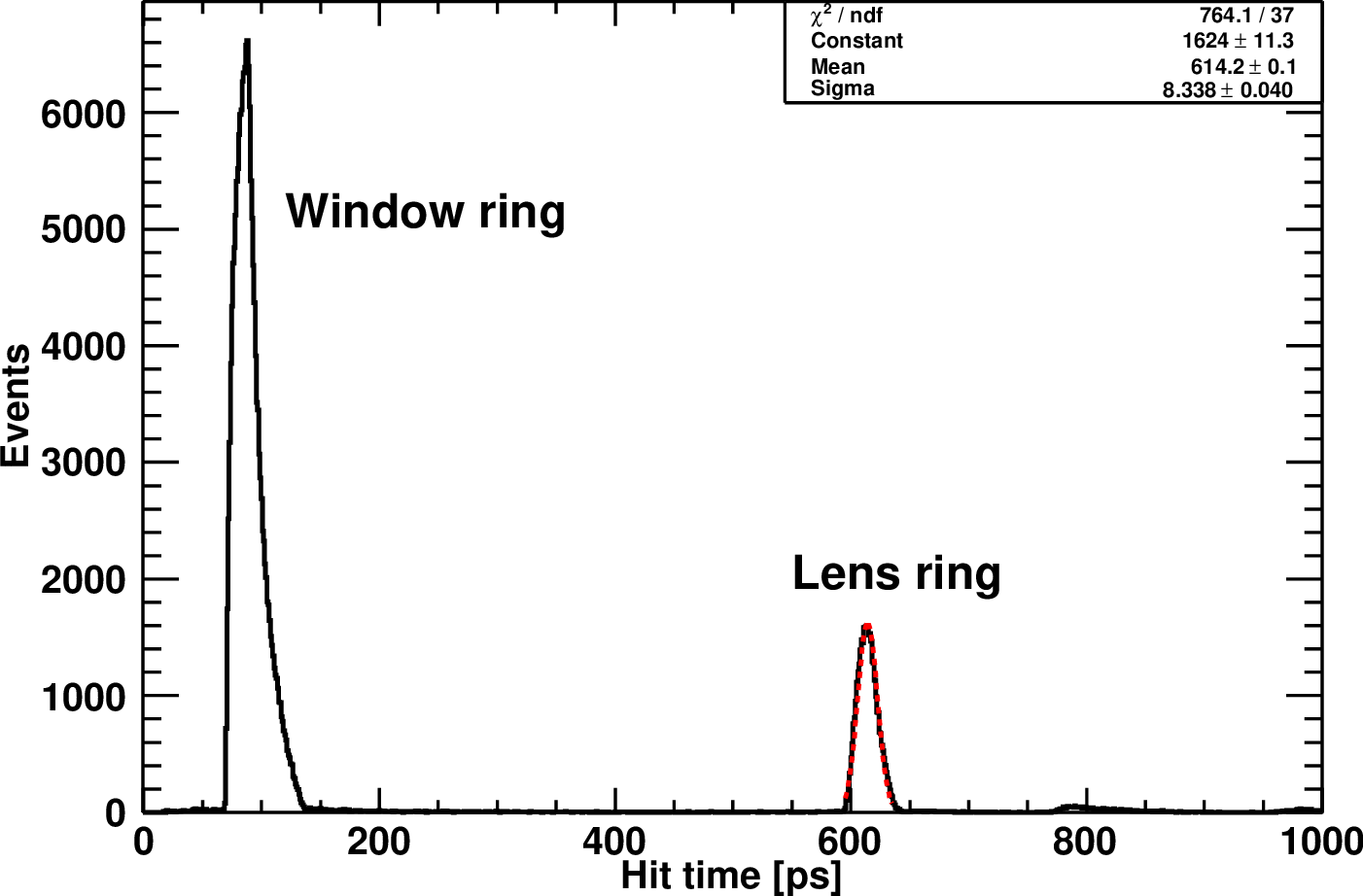}
\caption{\label{fig:geant4_r_t}Photon radius (left) and time (right) at the LAPPD photocathode obtained from Geant4 simulations. The red dashed curve  in the right plot is a gaussian fit of quartz lens Cherenkov photons.
}
\end{center}
\end{figure}

The distributions of a radius and time for the photons reaching the PC and successfully converted into PEs are shown in Fig.~\ref{fig:geant4_r_t}. The first peak in these distributions is due to the Cherenkov photons produced in the LAPPD window by the primary hadron. They are reflected 
back at the window-to-air boundary 
towards the PC, where part of them is converted into photo-electrons. The second, smaller peak is related to the photons of the Cherenkov ring produced in the radiator lens. The ring radius at the photocathode is located in the region of 57-67~mm, corresponding to the second row of pads from the beam spot one. In horizontal and vertical directions there is a $\sim$28\% chance for a photon to be detected in a third pad. The time of arrival of the lens Cherenkov photons is about 500~ps larger than for the direct photons from the LAPPD window. The timing resolution due to the geometry and chromatic dispersion is about 8.3~ps r.m.s.
%

\begin{figure}[!ht]
\begin{center}
\includegraphics[scale=0.25, angle=270]{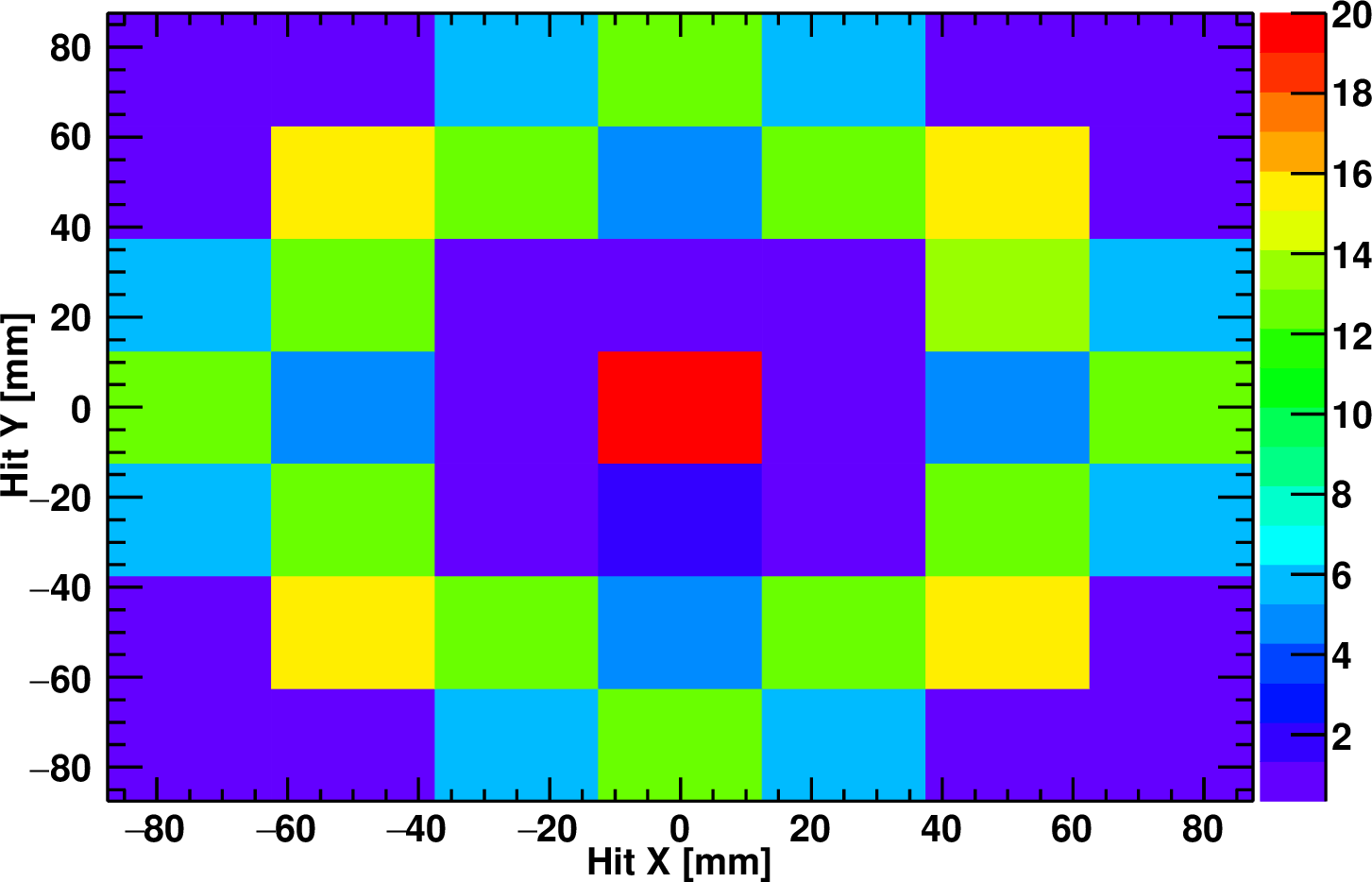}
\caption{\label{fig:geant4_npe_map_no_af}Same as in Fig.~\ref{fig:geant4_npe_map} (right), but simulated without the acrylic filter. Here, the maximum $Z$-value is set to 20.
}
\end{center}
\end{figure}

Without an acrylic filter the mean number of photo-electrons from the radiator lens per readout pad was 5.8 times higher and the timing resolution due to the geometry and chromatic dispersion was almost two times worse. Moreover, without the acrylic filter, the lens Cherenkov light was not contained in the second pad from the beam spot, but was extending to the third one, especially in the horizontal and vertical directions, where the maximum signal appeared in the third pad as shown in Fig.~\ref{fig:geant4_npe_map_no_af}.

%

\section{Test beam Setup}\label{sec:setup}
The experiment was performed at the CERN PS T10 test beam in the period of 5 -- 19 October 2022. The facility was shared with the ePIC dual RICH (dRICH) group, testing a detector prototype. Our measurement didn't require beam particle tracking, since simply selecting tracks in a narrow area after the radiator served our purpose. 
The LAPPD setup was installed downstream of the dRICH prototype.

\begin{figure}[!ht]
\begin{center}
\includegraphics[width=0.95\textwidth]{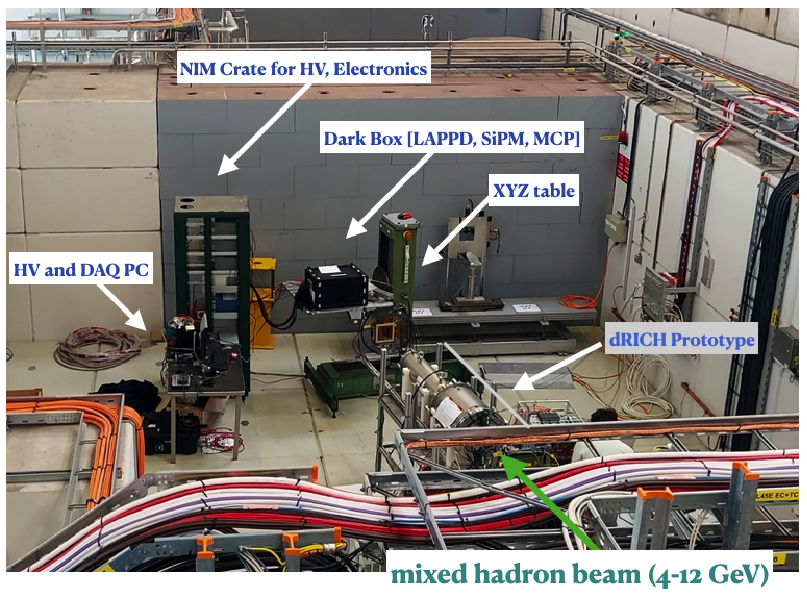}
\caption{\label{fig:T10_Hall} The LAPPD setup at the T10 beamline of CERN PS: the detector is located inside the black box near the wall.}
\end{center}
\end{figure}
\begin{figure}[!ht]
\begin{center}
\includegraphics[width=0.95\textwidth]{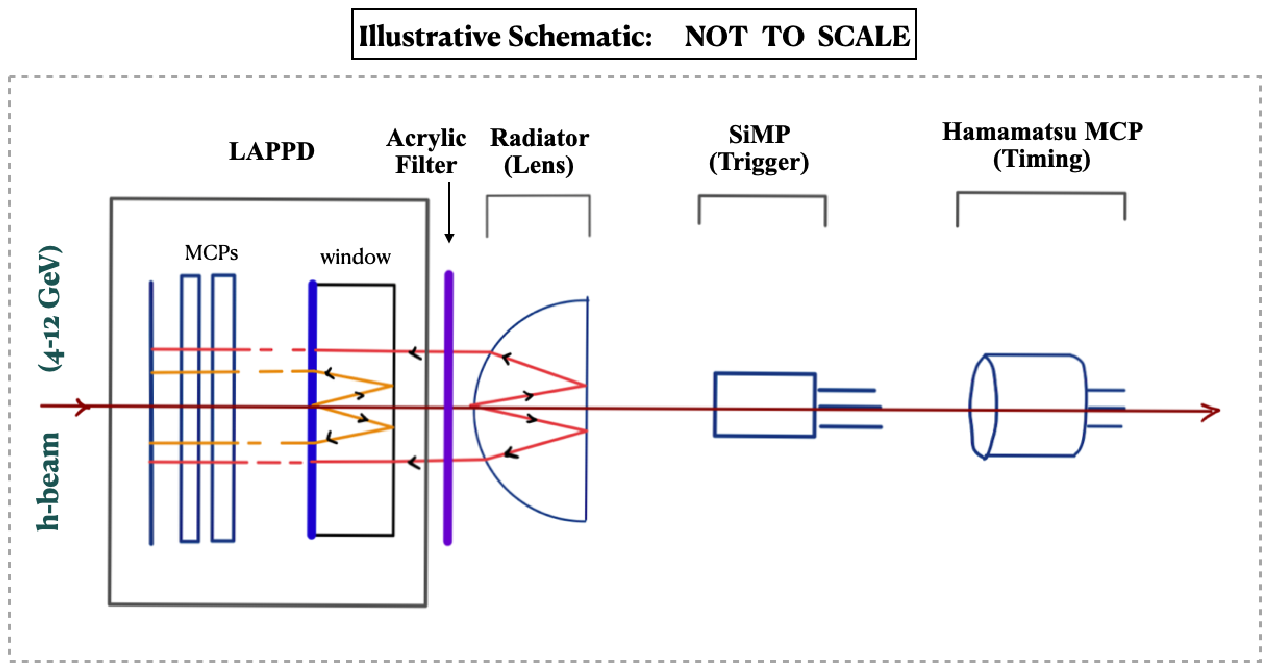}
\caption{\label{fig:inside-darkbox} The LAPPD setup inside the dark box.}
\end{center}
\end{figure}

%
In Fig.~\ref{fig:T10_Hall}, an overview of the T10 experimental hall as during our measurement is shown. The particle beam, indicated by a green arrow, contained a mixture of hadrons (pions, kaons and protons) with momenta selected in a range from 4 to 10 GeV/c. The LAPPD setup was enclosed within a dark-box, which was then placed on a movable table named $XYZ$-table. The $XYZ$-table allowed for 3-D movement with a sub-millimeter precision by a remote control, making it possible to align the setup with respect to the beam. 
Inside the dark box (Fig.~\ref{fig:inside-darkbox}), we mounted the LAPPD, followed by a plano-convex aspheric quartz lens as a radiator. 
\par
Further downstream, a bundle of scintillating fibers coupled to a Hamamatsu MPPC-SiPM provided triggering capabilities, while a Hamamatsu MCP-PMT was used as a timing reference (discussed in detail Sec.~\ref{sec:calib_meas}). The signal and supply cables passed through the dark box in a light tight arrangement. 
%
%
\par
A rack with setup services was placed near
the LAPPD setup. It hosted a NIM crate, High Voltage (HV) and low voltage power supply units and a DAQ dock. A Linux PC running the DAQ system had been installed on a nearby table. 

Signals from the read-out pads were taken from the SMA connectors mounted on the readout PCB. All unused SMA connectors were shorted to ground.
We used four eight-channel custom made inverting amplifiers (discussed in detail in Sec.~\ref{sec:readout_elec}) for signal amplification. Signals from all the detectors (LAPPD, SiPM, Hamamatsu MCP) were acquired by a CAEN V1742 digitizer board. The SiPM signals were connected to TR0 and TR1 fast trigger inputs of the V1742 module, and were digitized along with the LAPPD and MCP-PMT ones. The signals from the Hamamatsu MCP-PMT and 31 LAPPD pads of LAPPD were read-out by the 32 input channels of the digitizer. The 31 active pads of the LAPPD were selected in the region where signals produced by the radiator light cone and the beam signal were expected. 


%
%

The secondary beam in T10 experimental area was relatively wide, with a transverse size being of the order of 1.5~cm$^2$. Preliminary Geant4~\cite{geant4} simulations indicated that such a large beam spot, limited for our measurements only by the 1~cm diameter of the Hamamatsu MCP-PMT, would lead to a 15~ps uncertainty in the Cherenkov photon timing. In order to reduce this beam size uncertainty, a beam monitor detector was built. The detector was made of a 5$\times$5 mm$^2$ bundle of scintillating fibers Kuraray\footnote{Kuraray Plastics Co.,Ltd., Chiyoda City, Tokyo, Japan} 3HF(1500)MJ. Each single fiber had a diameter of 0.5~mm and a total length of 10~cm. The bundle had a 5~cm straight section aligned along the beam followed by a 90$^o$ turn to allow coupling of the fibers to a Hamamatsu MPPC S13360-6025CS (referred to as a SiPM in this article) of 6$\times$6 mm$^2$ size. The SiPM signal was amplified with a transimpedance amplifier~\cite{hps_amplifier} with two identical outputs, used, as mentioned earlier
, to feed the V1742 digitizer fast trigger inputs TR0 and TR1. Given the length of the straight section of the beam monitor, we expected about 600~PE signal in the SiPM per beam particle. The rise time of the SiPM amplifier was $\sim$7~ns and the signal-to-noise ratio of the order of 350.  However, the observed coincidence time resolution with the Hamamatsu MCP-PMT was never better than 150~ps, obtained at the highest signal amplitude.
\section{Readout Electronics and data acquisition}\label{sec:readout_elec}
The main components of the readout electronics, discussed in the following, are the custom amplifiers used for the amplification of the PE signals generated in the LAPPD and in the MCP-PMT and the commercial digitizer V1742 by CAEN.
\subsection{The custom amplifier}\label{sec:amplifier}
The LAPPD signals produced by single photoelectrons have a typical 
amplitude around 20~mV. Therefore, in order to explore the full performance of the LAPPD timing capabilities, its output signals were amplified with a custom broadband amplifier.

A Microwave Monolithic Integrated Circuit (MMIC) Darlington
amplifier has been used. It has a high dynamic range, a wide analog bandwidth of 2~GHz, a gain of 20~dB, input and output impedance internally matched to 50~Ω.
The MMIC package SOT-89 looks like a discrete transistor, fabricated using InGaP HBT technology (Indium-Gallium-Phosphide Heterojunction Bipolar Transistors).

A Darlington amplifier is a 2-port device: RF input, and combined RF output and bias input.
It is housed in a 4-lead package including 2 ground leads; when both of them are connected to the external
ground, the common path impedance is minimized  allowing for the best RF performance. The internal resistors 
determine the DC operating point of the transistors and provide feedback to set RF
gain, bandwidth, and input and output impedance to optimal values.

%
%

\begin{figure}[!ht]
\begin{center}
\includegraphics[width=0.5\textwidth]{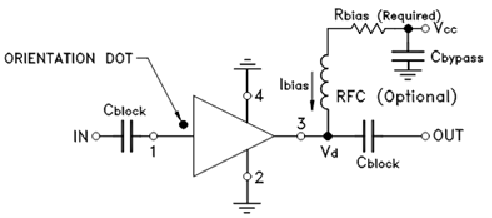}
\caption{\label{fig:Biasing}Biasing configuration for Darlington amplifier.}
\end{center}
\end{figure}

A MMIC biasing configuration is shown in Fig.~\ref{fig:Biasing}. The bias current is delivered from a voltage
supply
$V_{cc}$ through the resistor $R_{bias}$ and an RF choke (inductor, not mounted in our application), shown as RFC in
the figure. The resistor reduces the effect of a device voltage ($V_d$) variation on the bias current by
approximating a current source. Blocking capacitors are needed at the input and output ports. They should have
low Effective Series Resistance (ESR) and should have a low enough reactance to exclude the negative effects on the insertion loss and Voltage Standing Wave Ratio (VSWR) at low frequency. The blocking capacitors must be free of parasitic resonances
up to the highest operating frequency. The use of a bypass capacitor at the $V_{cc}$ end of $R_{bias}$ is advised to prevent a stray coupling from or to other signal processing components via the DC supply line.

A PCB developed for the test beam, hosts eight identical channels, and has a two-layer structure, made of
Rogers RO4350 material with relative dielectric constant 3.5, thickness 0.8~mm, with controlled 
50~$\Omega$ impedance traces.
A connection to inputs and outputs is made via SMA-type RF connectors soldered directly to the PCB traces to
minimize the external noise pick-ups and 
a cross-talk with the neighboring channels. Low-voltage power was provided via a
bus-bar distribution. The whole system was placed inside a shielding metal enclosure, used also to
dissipate heat produced by the amplifiers, which require about 600~mW/ch.
The V1742 digitizer input noise was measured to be $\sim$0.5~mV; therefore, a use of the amplifier increased the signal-to-noise ratio from 13 to $\sim$500. Combining this value with the LAPPD signal rise time, which is about 0.7~ns,  we estimated that the
contribution of the electronics readout chain to the timing resolution was $<$5~ps.

\subsection{The V1742 digitizer}\label{sec:digitizer}
\par
CAEN V1742 digitizer VME module~\cite{v1742} is based on the DRS4 (Domino Ring Sample) 8-channel chip~\cite{DRS, DRS4,DRS4_old}. Four DRS4 chips are hosted in a V1742 digitizer, resulting in  32 analog channels (0-31) and two fast, namely low latency trigger channels (TR0 and TR1) as inputs; each of the fast trigger is split to serve two DRS4 chips. The input channel dynamic range is 1~V peak-to-peak.
The DRS4 is a Switched Capacitor Array (SCA). Each channel is equipped with an array of 1024 capacitors used to operate in a sample and hold mode. Inverter chains as delay lines boost the sampling speed into the giga-samples per second range. This continuous sampling is controlled by a Domino Wave circuit, which makes this array acting as a circular buffer.
On average, at 5~GS/s, the delay between the samples is $\sim$200~ps. However, the delay between the samples is largely non-uniform. 
Fine timing information can be obtained only after a careful calibration of all the 1024 sampling cells of each channel, as described in Sec.~\ref{sec:calibration}. The (32 + 2) channels of the digitizer used at the test beam have been calibrated.
\par
The analog to digital conversion is not simultaneous with the chip sampling phase, as it starts as soon as a trigger condition is met. When the trigger stops, the DRS4 chip sampling (holding phase), the analog memory buffer is frozen, and the cell content is made available to a 12 bit ADC for the digital conversion. Different trigger sources are available. For data taking, we have used fast triggers TR0 and TR1, which are  convenient for high precision timing measurements, since signals to TR0 and TR1 can be digitized, reported in the output data and used as a time reference.  In the calibration procedure both these trigger inputs and the slower global trigger have been used.
\subsubsection{Digitizer calibration}\label{sec:calibration}
We have used a 33600A waveform generator by Keyseight\footnote{Keyseight, 1400 Fountaingrove Parkway Santa Rosa, CA 95403-1738, USA}. Identical calibration procedures were independently applied for each channel. The calibration of a single channel is reported as an example. 
\paragraph{Amplitude Calibration}
This calibration allows us to obtain, for each of the 1024 capacitor cells of a channel,  the parameters needed for an ADC to voltage conversion. Three constant DC voltages (400 mV, 0 V, -400 mV) were used. An uncorrelated trigger signal is used to obtain the holding and read-out phase. This signal  feeds the fast trigger channels when calibrating channels 0-31 and the global trigger to calibrate channels TR0 and TR1.   A linear fit is performed to extract the amplitude calibration parameter for each of the 1024 cells.
\par
The noise resolution of the DRS4 chip in use has been determined using the calibration parameters. The resulting noise is $\sim$0.5~mV.
\paragraph{Time Calibration}
    The detailed description of the time calibration method we have applied can be found in ref.~\cite{ritt_calibration}.
    \par
    The accurate Timing Calibration (TC) requires the amplitude calibration parameters. During the calibration procedure, a channel not used in the TC data taking had always been used to monitor the baseline. This guarding channel allows us to detect and remove, when needed,  any unwanted spike affecting the calibration measurements. A trigger similar to the ADC calibration was used. The TC proceeds in two steps. First, a  TC is performed calibrating all the cells with respect to a single one arbitrary selected. Then, a global TC of all the cells is performed by taking as a reference the already calibrated ones. 
    \par
    A sinusoidal signal at 50-100~MHz is used with a precisely known frequency and a fast signal trigger is injected in the fast trigger inputs. At a zero crossing, a liner interpolation of the rising and of the falling edge is performed. Time width of each cell is determined by taking the difference of subsequent zero crossing points. In order to cancel some residual effects of the amplitude calibration, an average of the rising and falling edges is applied. 
    \par
\begin{figure} [!ht]
    
   \begin{minipage}{0.5\textwidth}
	\includegraphics[trim=8cm 3cm 1cm 1.5cm, clip,width=1\linewidth]{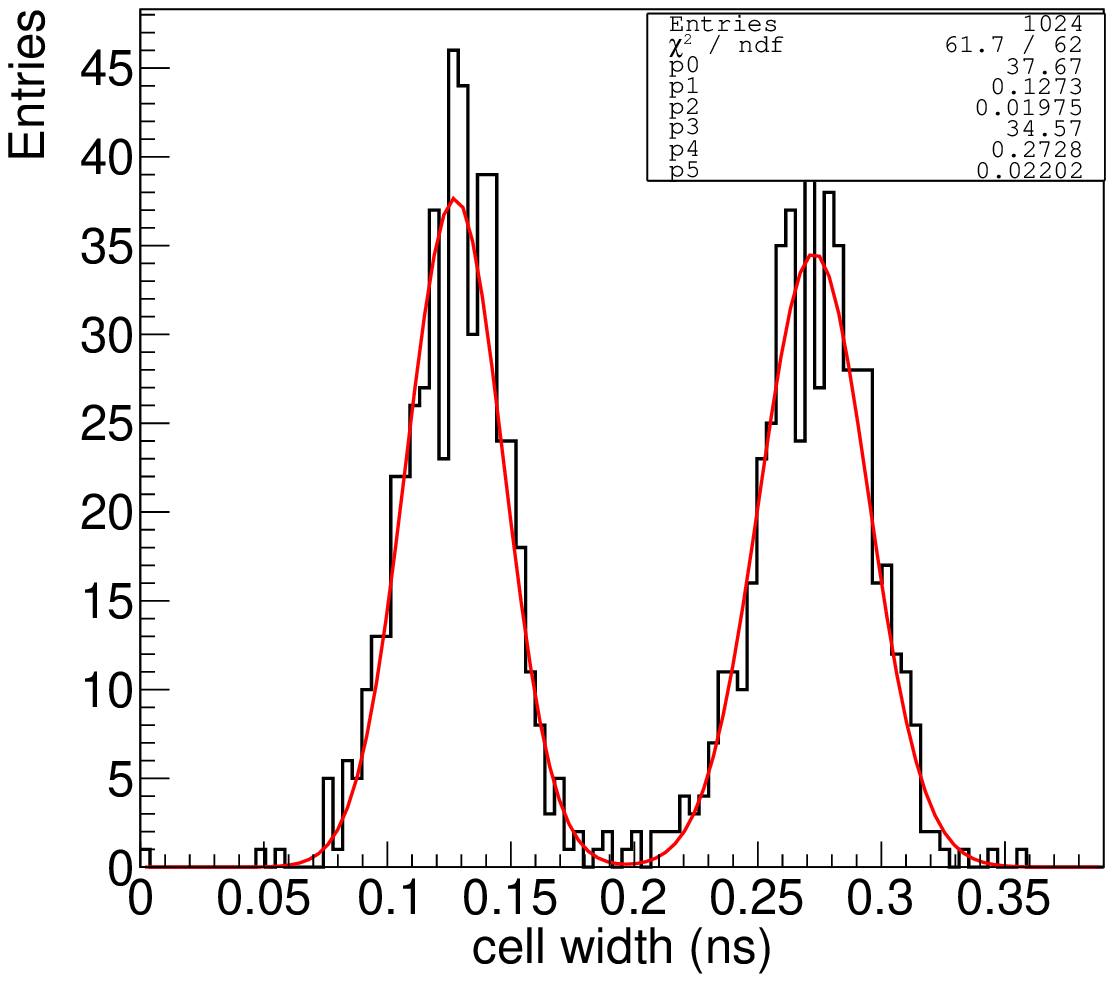}
    \end{minipage}%
    \begin{minipage}{0.5\textwidth}
	\includegraphics[trim=8cm 2.5cm 1.5cm 1cm,clip,width=1\linewidth]{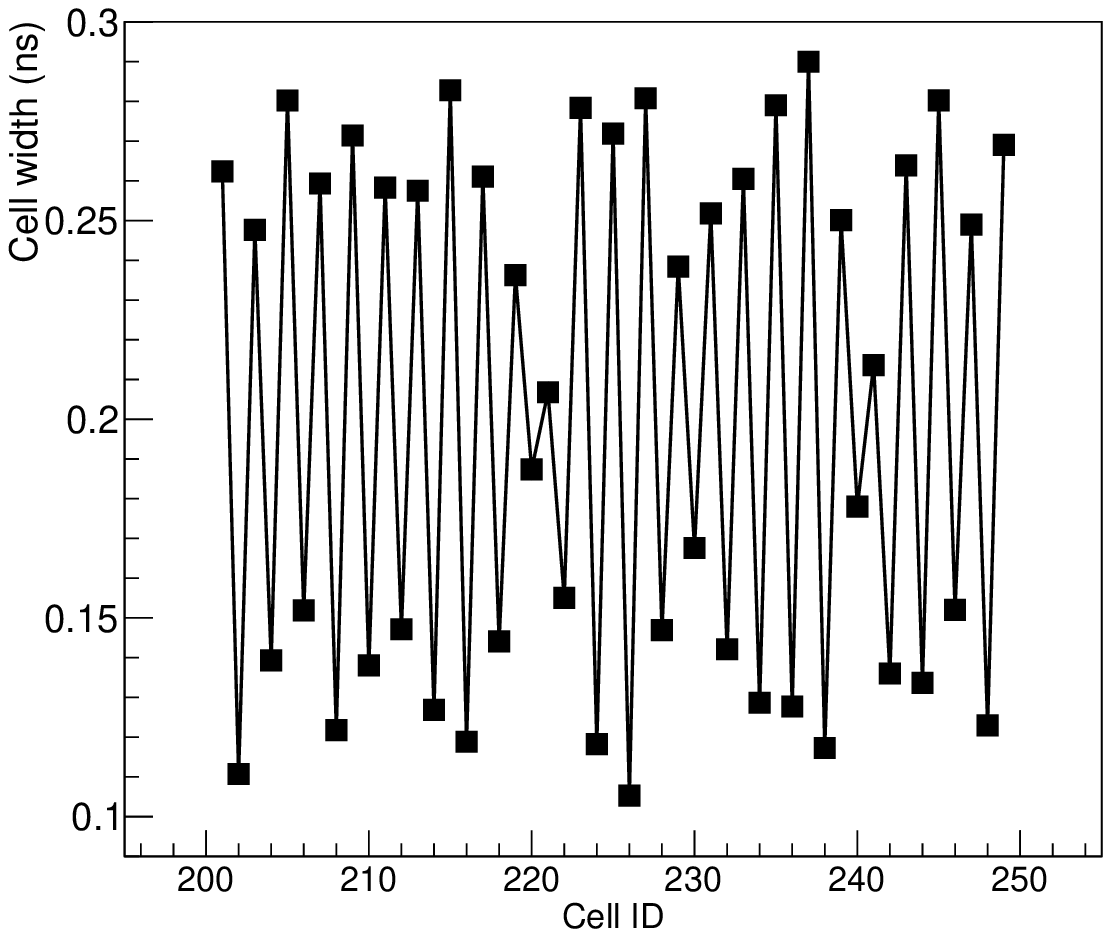}
    \end{minipage}
    \caption{\label{fig:TC}Left: Time width of different capacitor cells. Right: An example of time width versus channels for 50 cells: a systematic dependency of time width with odd-even cells is observed.}
\end{figure}

    It has been observed (Fig.~\ref{fig:TC}) that the cell time widths vary substantially for even and odd cells and each cell may have more than 50 ps differences from a nominal 200 ps width.  The results are compatible with what was reported in ref.~\cite{kimetal}.
\paragraph{Validation }
The calibration outcome has been validated for each channel. A pulse with a 3 ns rise time was split and fed into two channels of the digitizer. The signal time is defined as a 50\% amplitude crossing. The r.m.s. of timing difference is assumed as the time resolution achieved using the calibration. The resolution is $\sim$2~ps for eight channels of the same DRS4 chip and $\sim$3~ps for channels of different DRS4 chips. When one of the two signals is delayed by 50~ns, the resolution obtained is $\sim$4~ps.

\begin{figure}[!ht]
\begin{center}
\includegraphics[scale=0.25, angle=270]{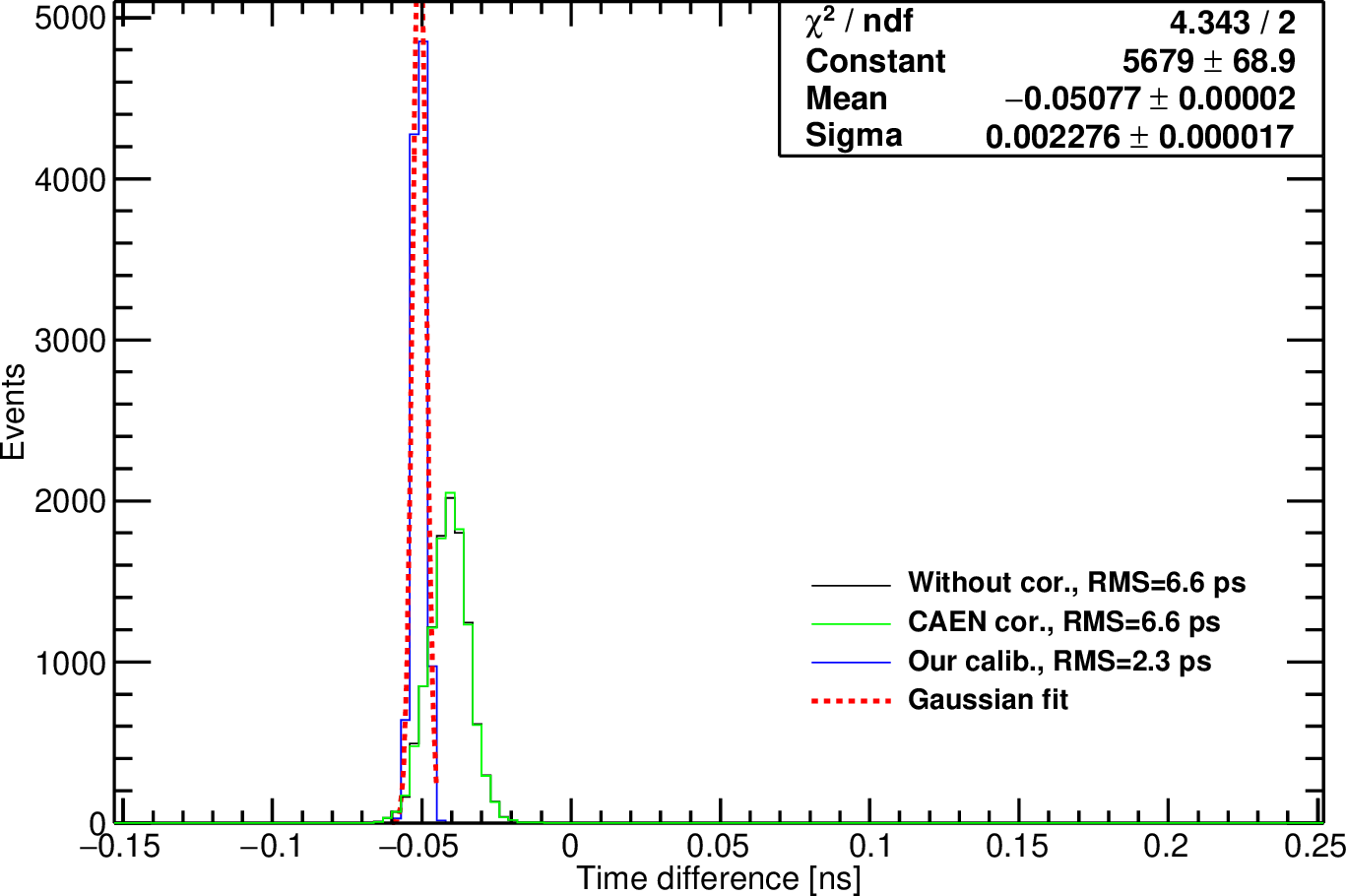}~%
\includegraphics[scale=0.25, angle=270]{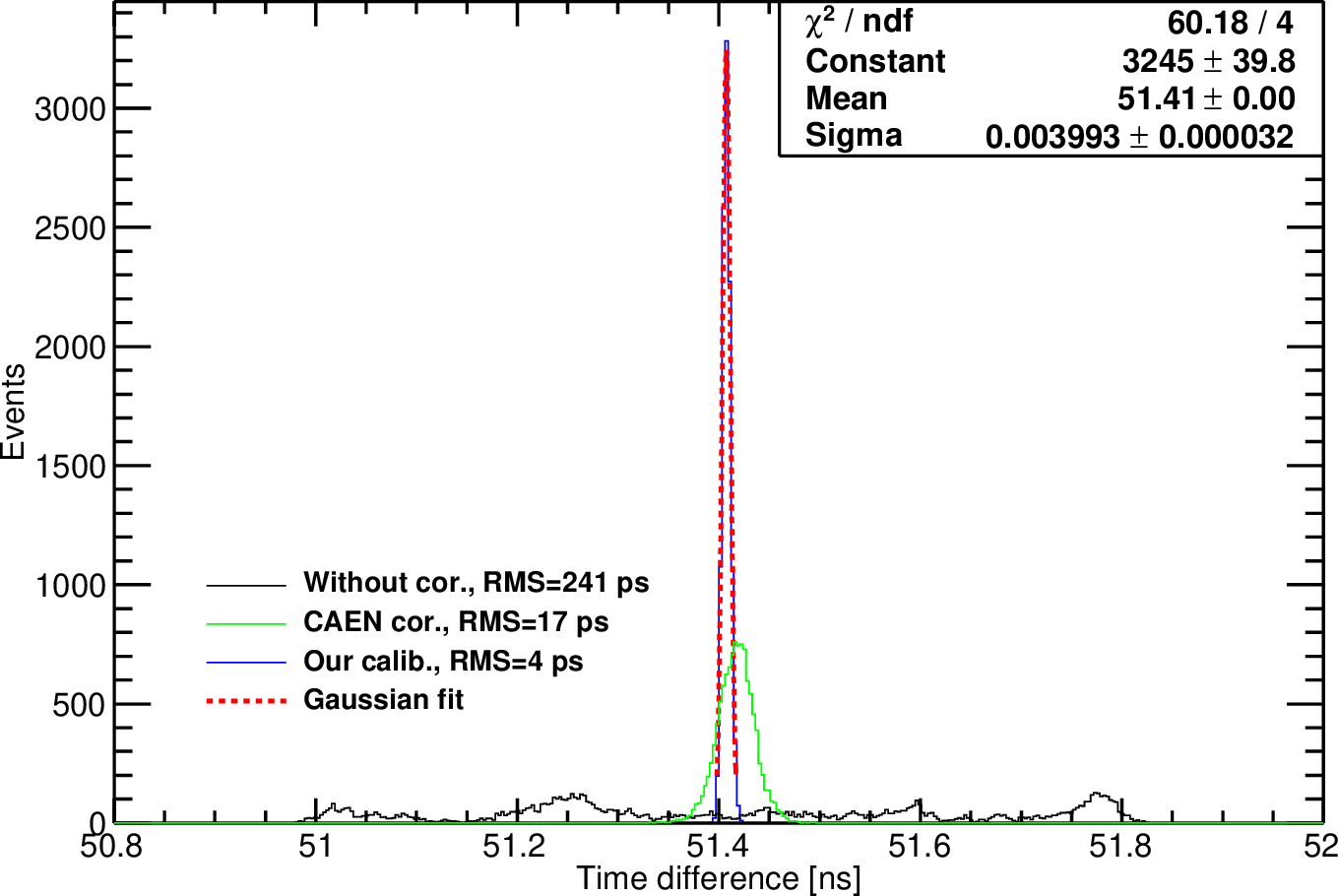}
\caption{\label{fig:v1742_t_calib}Example of a validation of the V1742 timing calibrations: time difference between channel 1 and channel 0 for a zero delay (left)
and for a delay of 51~ns (right). The distributions obtained without applying any correction (black), using the corrections from CAEN table (green, on the left plot coincides with black) and by the calibration performed by us (blue) are overlapped, together with a Gaussian best fit to this last distribution (dashed red).}
\end{center}
\end{figure}

\paragraph{Timing calibration results}
The resulted delay tables were validated to give $<$5~ps timing resolution (Fig.~\ref{fig:v1742_t_calib}). The calibrations provided by CAEN on the V1742 flash memory also improve timing resolution, especially for a large time difference, but remain significantly worse than the results of the procedure from Ref.~\cite{drs4_time_calibration}.

\subsection{The data acquisition}\label{sec:DAQ}
The two beam monitor signals used to trigger the digitizer through the TR0 and TR1
fast trigger inputs were digitized together with the LAPPD pad and Hamamatsu MCP-PMT signals. They provided a common reference to the four DRS4 chips inside V1742. Using CAEN WaveDump software the data were saved to disk through a VME USB controller V1718 by CAEN. An Ubuntu linux PC was used to run the DAQ and store the data.
The average rate was about 20~Hz, however during the 0.4~s spill the instantaneous rate reached 2~kHz.

\section{Single photoelectron Amplitude Calibrations}\label{sec:calib_meas}
The timing measurements of interest for applications in Cherenkov detectors should be performed on Single Photo-Electron (SPE) signals. Therefore, the measurements of the SPE signal amplitude defined the working range, where the timing studies were performed.
\par
\begin{figure}[!ht]
\begin{center}
\includegraphics[scale=0.25, angle=270]{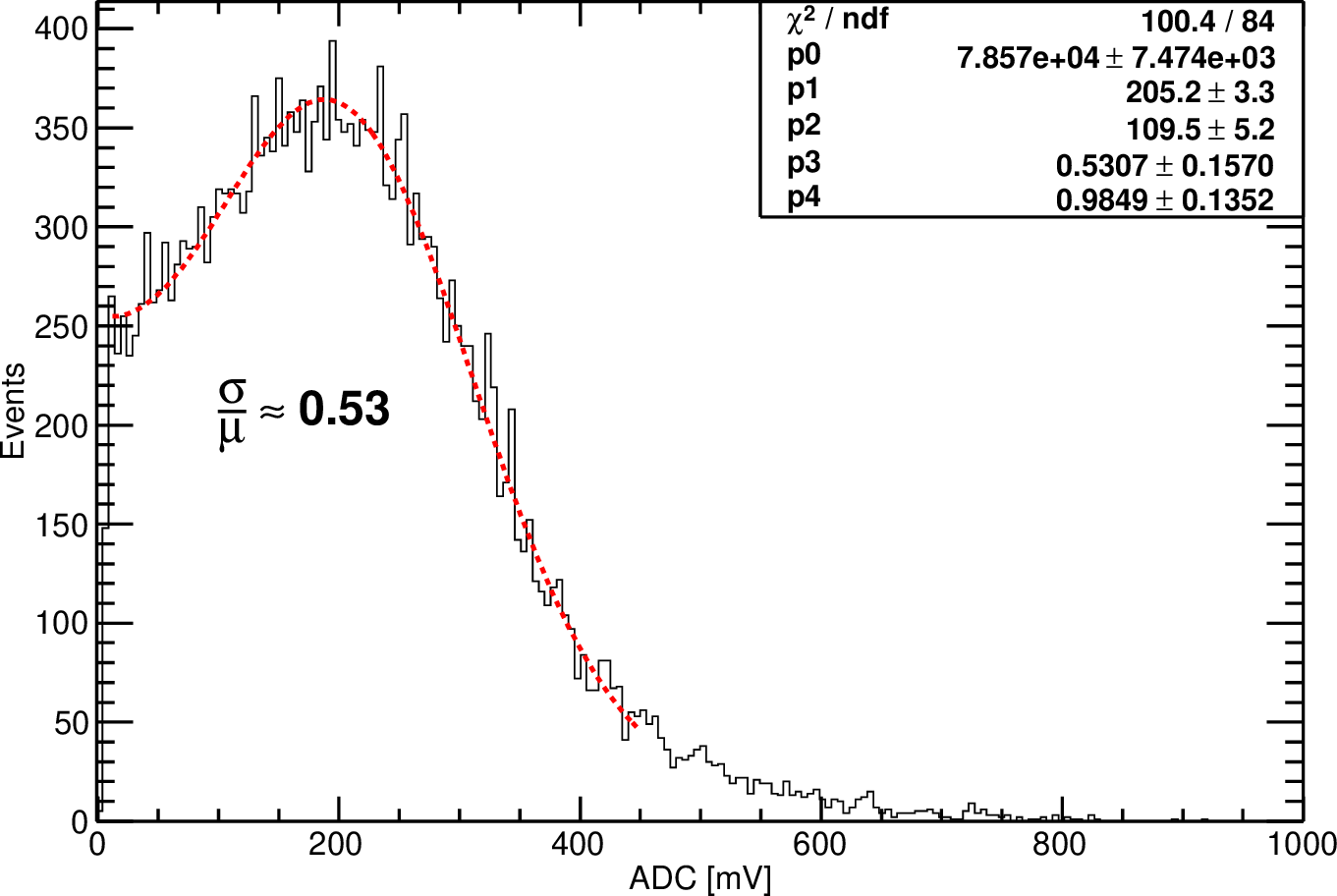}~%
\includegraphics[scale=0.25, angle=270]{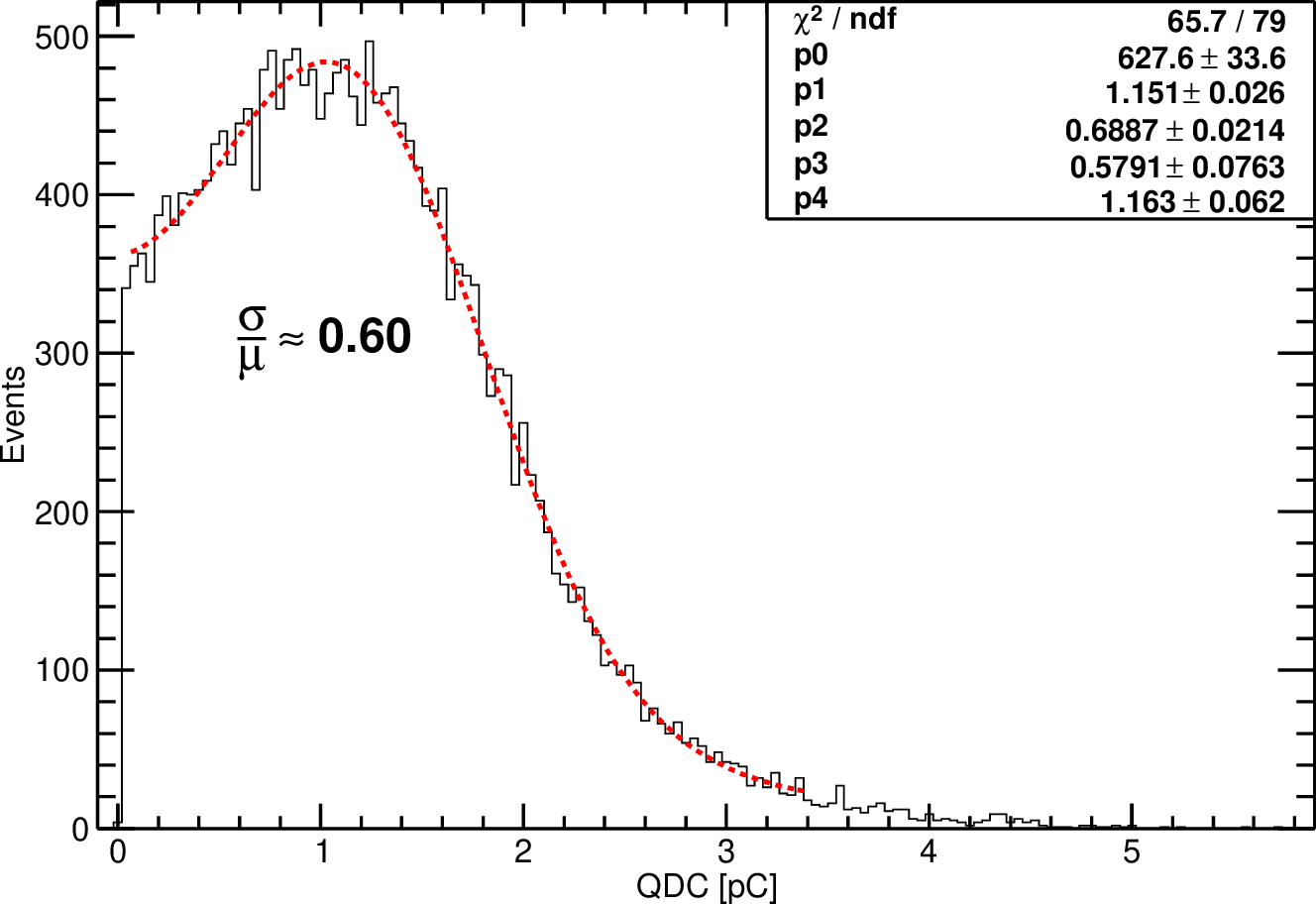}
\caption{\label{fig:phd_qdc_spe_fit} Pulse height and collected charge distributions of SPE signals measured with the digitizer. The dashed curves show the fit to the data with a convolution of an exponential and Gaussian distribution.}
\end{center}
\end{figure}

The charge collected on the LAPPD anode induces 
a fast leading signal on readout pads, followed by a long discharge tail. 
This fast leading signal was measured with the digitizer and its integral in a 3~ns time interval, properly converted from amplitude to charge information,
was used to estimate the collected charge.
This measurement was cross-checked by using an oscilloscope with larger analog bandwidth allowing to find the pulse height and to integrate the signal over its duration. Making use of the digitizer information, both the integrated charge and the pulse amplitude can be used as estimators of the collected charge. The two measurements exhibit similar precision. In fact, the measured pulse height and integral distributions  shown in Fig.~\ref{fig:phd_qdc_spe_fit} exhibit a similar relative resolution for the SPE peak position, which match within 10\%. The integral distribution also allows a direct evaluation of the absolute value of the collected charge. However, the pulse height measured with the digitizer is much easier to evaluate and provides a direct connection to the threshold and the dynamic range of the ADC. Thus, in the following text, the pulse height is mostly used as the estimator of the collected charge.

\begin{figure}[!ht]
\begin{center}
\includegraphics[scale=0.25, angle=270]{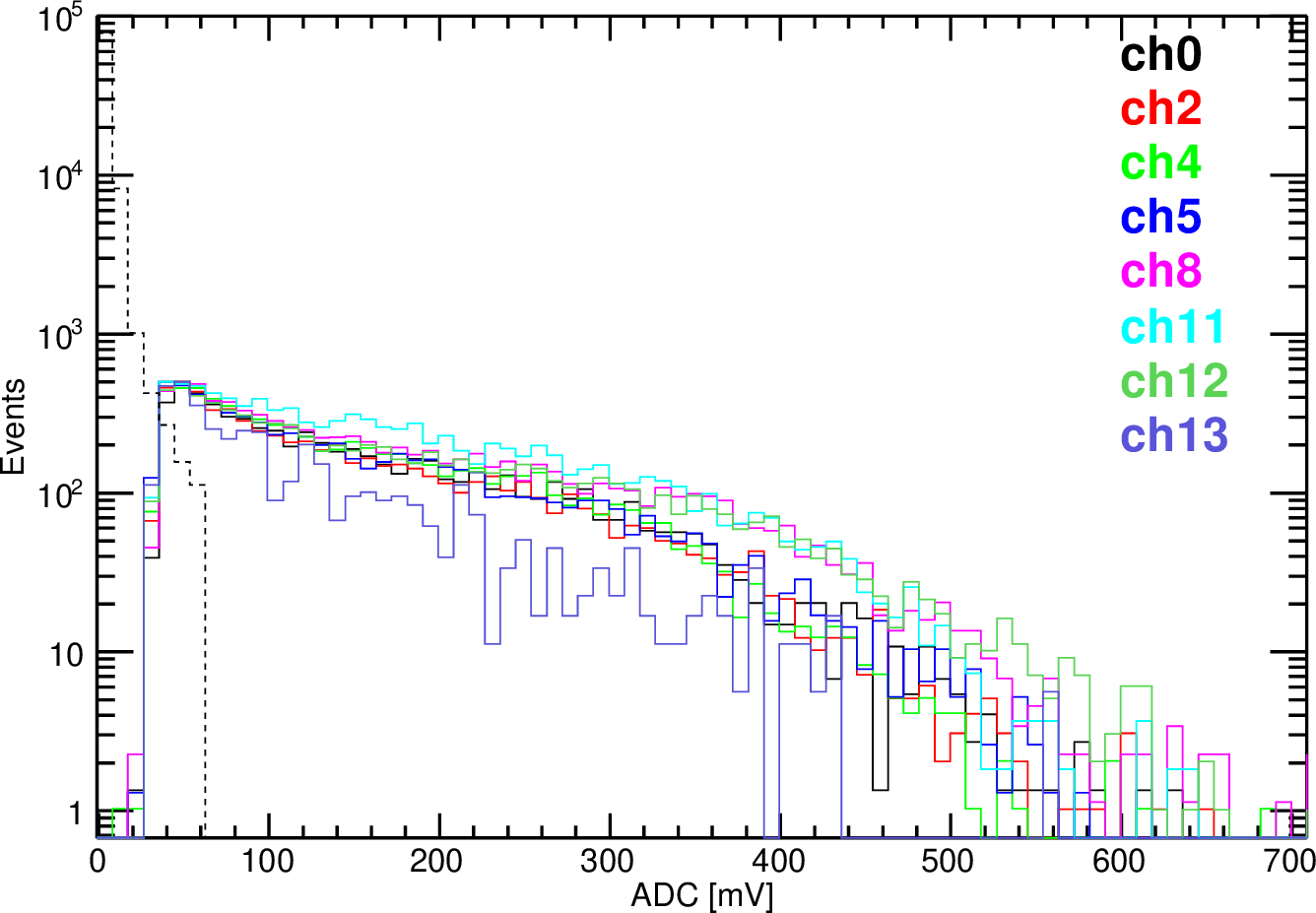}~%
\includegraphics[scale=0.25, angle=270]{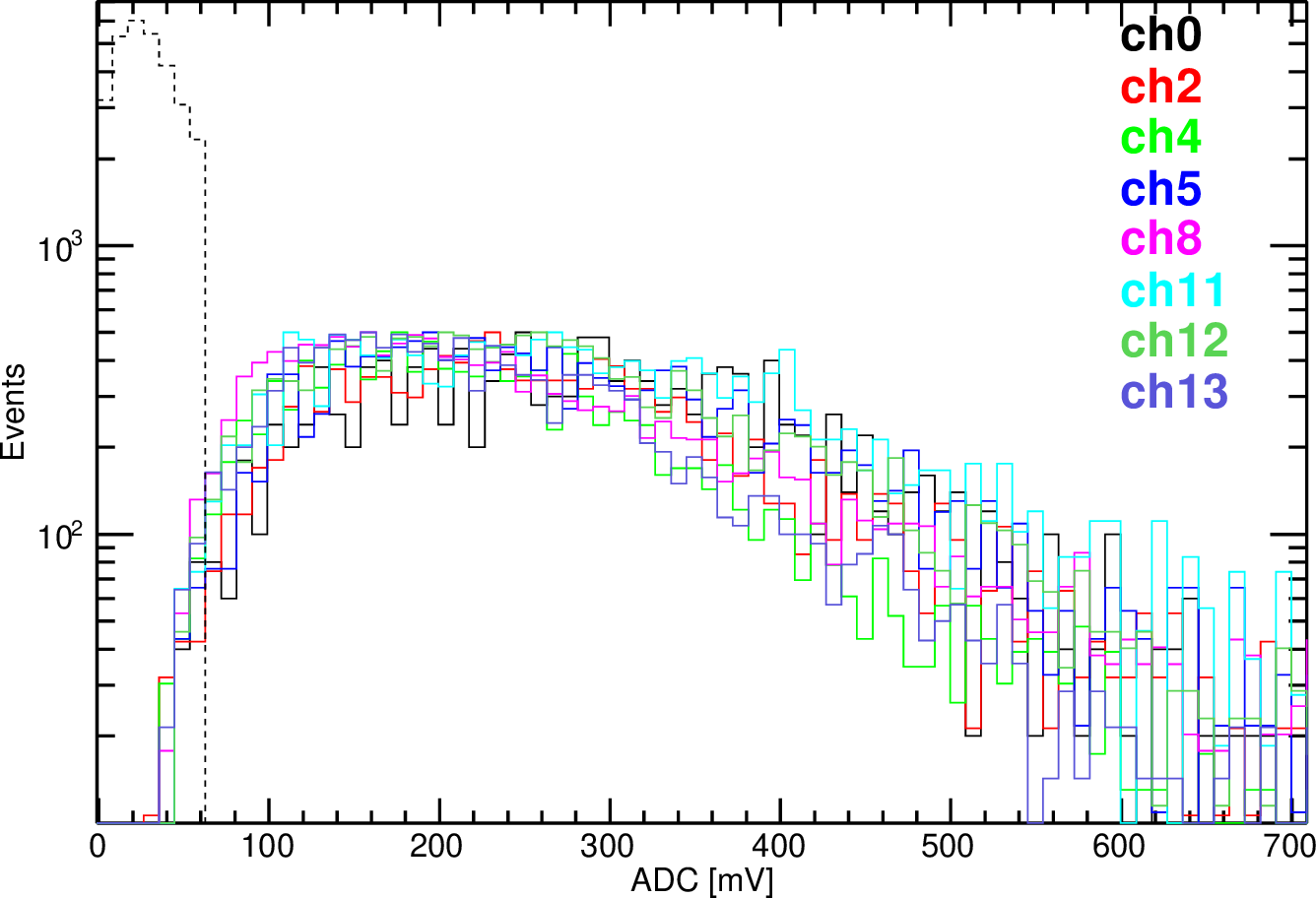}
\caption{\label{fig:phd_spe}Pulse height distributions of SPE signals measured in a LED calibration run (left) and during beam irradiation at CERN PS (right). Different colors show different LAPPD pads selected in the quartz lens Cherenkov cone region. Dashed line shows the corresponding pedestal distributions.}
\end{center}
\end{figure}

One SPE amplitude calibration was performed at CERN PS during the period of the data taking.
This procedure used a green LED pulsed with a low 1.2~V amplitude,
50~ns wide signals generated with a 200~Hz frequency.
Because of the long LED response time, its emission interval was considerably longer than 50~ns (of the order of 1~$\mu$s). However, we selected only
the first signal from the trigger, taken from the generator sync output,
withing 40~ns from the main peak in the overall time distribution.
Trailing photo-electron signals, following the first one, were ignored.
In each individual channel, the fraction of events with a photo-electron signal in the selected time gate was
found to be about 9\%. Therefore, the measured distributions
were dominated by SPE contribution and the contamination of multiple photo-electron emission
was less than 4\%.
The pulse heights of these signals were saved and are shown in Fig.~\ref{fig:phd_spe}.
The uncorrelated distributions featuring pedestals were obtained from the digitizer data in a fixed interval after the synchronous trigger output of the generator, regardless the presence of the SPE signal.
\par
\begin{figure}[!ht]
\begin{center}
\includegraphics[scale=0.25, angle=270]{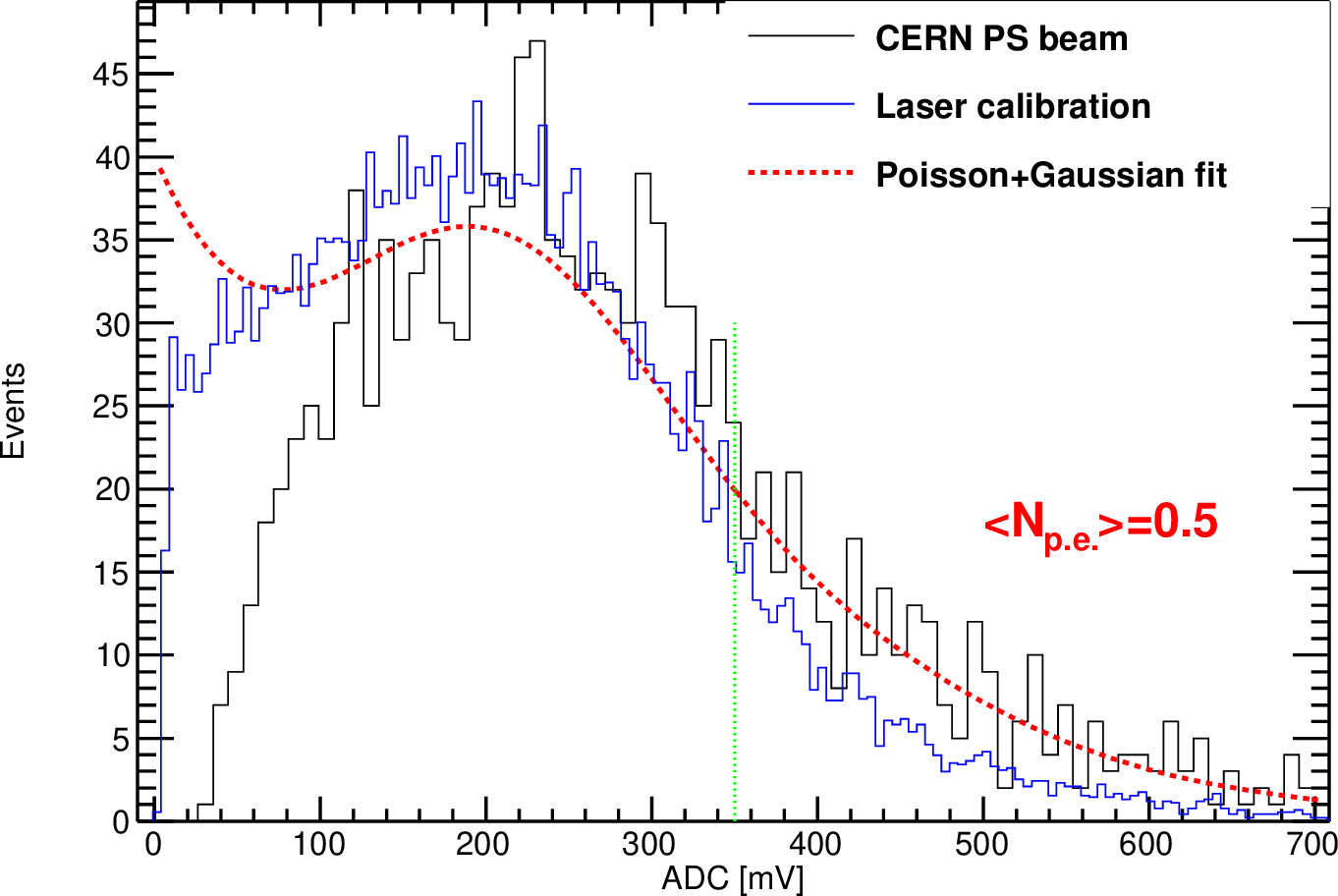}~%
\includegraphics[scale=0.25, angle=270]{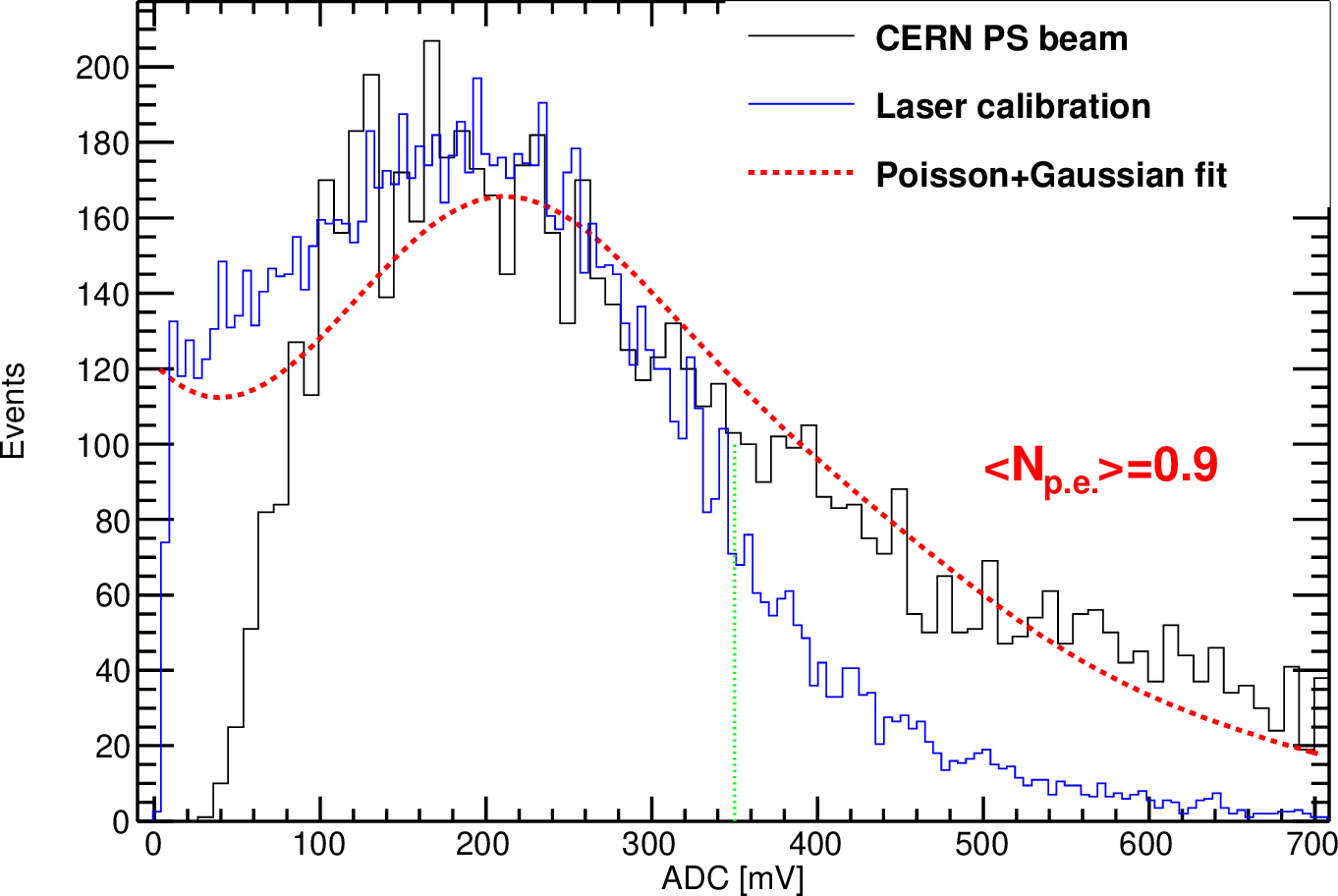}
\caption{\label{fig:cmp_phd_spe}Amplitude spectra measured in the Cherenkov ring readout pad F6 (left) and in the beam spot pad D5 (right) from testbeam data (black) compared with the laser calibration distribution (blue).
The fit to the testbeam data with a convolution of Poisson and Gaussian distributions is shown by the red dashed line. The mean number of PE is provided by the fitting procedure. The green dotted vertical line indicates the cut applied to select SPE events. As evident in the spectra, 
the software threshold used for the testbeam data is higher, to suppress the contribution of the cross-talk (Sec.~\ref{sec:x_talk}).}
\end{center}
\end{figure}

A second amplitude calibration was performed in the laboratory. A pulsed laser source was used for illuminating a single LAPPD pad at a 200 Hz frequency. The light source was a picosecond laser-head LHD-P-C-405 controlled by a PLD 800-D single channel driver by PicoQuant\footnote{PicoQuant, Rudower Chaussee 29 (IGZ), 12489 Berlin, Germany}. The average emission wavelength was 405~nm and the source was operated at an average minimum power of 800~$\mu$W. The pulse duration was below 50~ps (FWHM). The diaphragm in front of the light source was regulated to obtain $\sim$3\% non-empty events to ensure a robust domination of SPE signals.
The examples of the spectra collected in this calibration are shown in Figs.~\ref{fig:phd_spe} and \ref{fig:cmp_phd_spe}, where they are compared with the spectra collected at the test beam.
From the charge spectra, we estimated the LAPPD gain being $6.9\pm 0.7\times 10^6$ when operated at the typical biasing voltage, reported in Fig.~\ref{fig:HV_schematic}. This gain is in agreement with Incom data sheets.

We compared the amplitude spectra collected at the test beam in the readout pads detecting the PE from the Cherenkov ring with those obtained from LED and laser calibrations in order to verify that these readout pads were measuring mostly SPE signals.
Similar comparison was performed for the beam spot pads to check that those had seen multiple photo-electrons.
These comparisons, shown in Fig.~\ref{fig:cmp_phd_spe}, allowed us to conclude that, during the test beam, the Cherenkov ring pads were measuring mostly SPE signals (the contribution of many photo-electron events $<$9\%), while the beam spot pads had at least a factor of two larger contribution by multiple PE events.

\begin{figure}[!ht]
\begin{center}
\includegraphics[scale=0.25, angle=270]{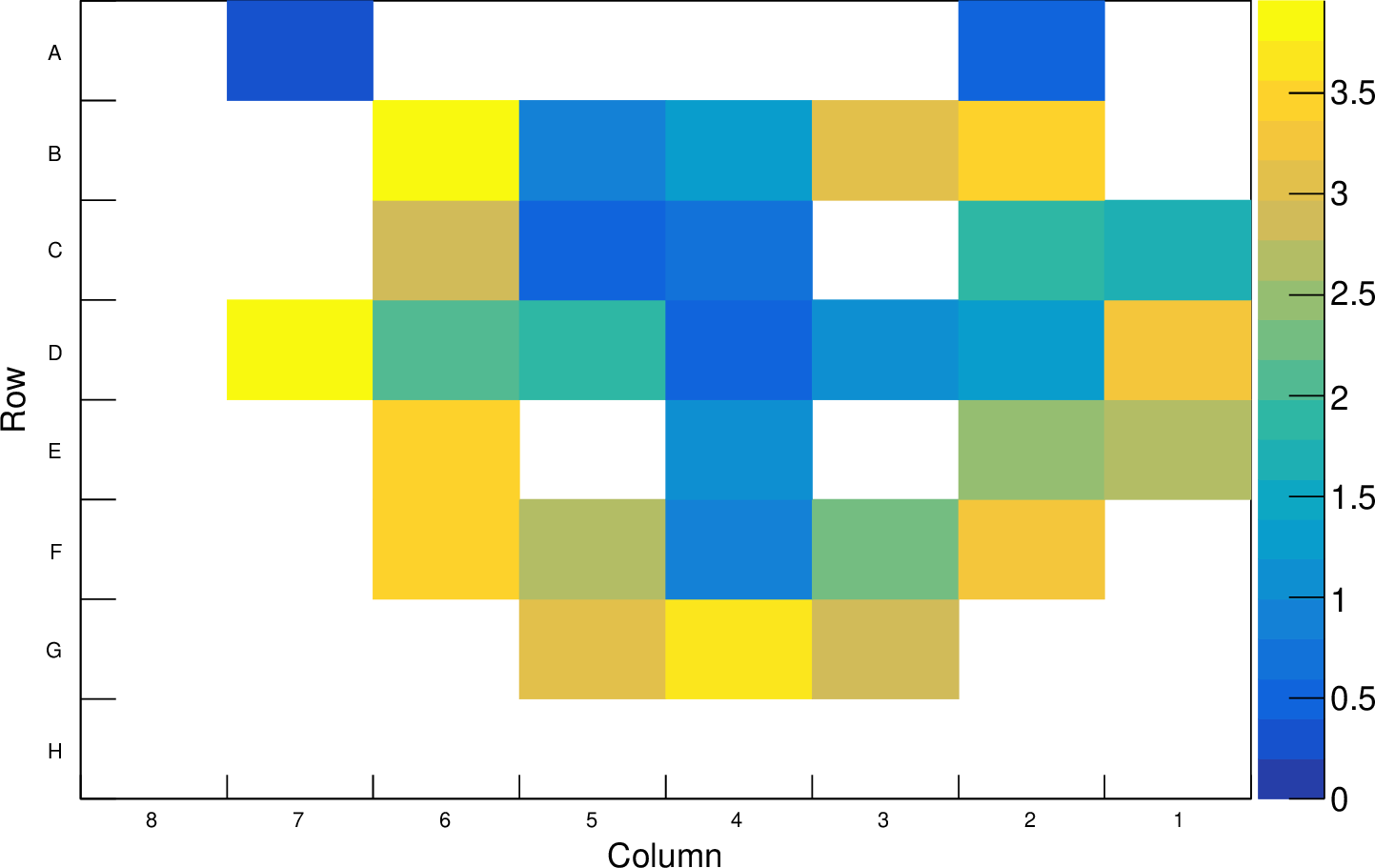}~%
\includegraphics[scale=0.25, angle=270]{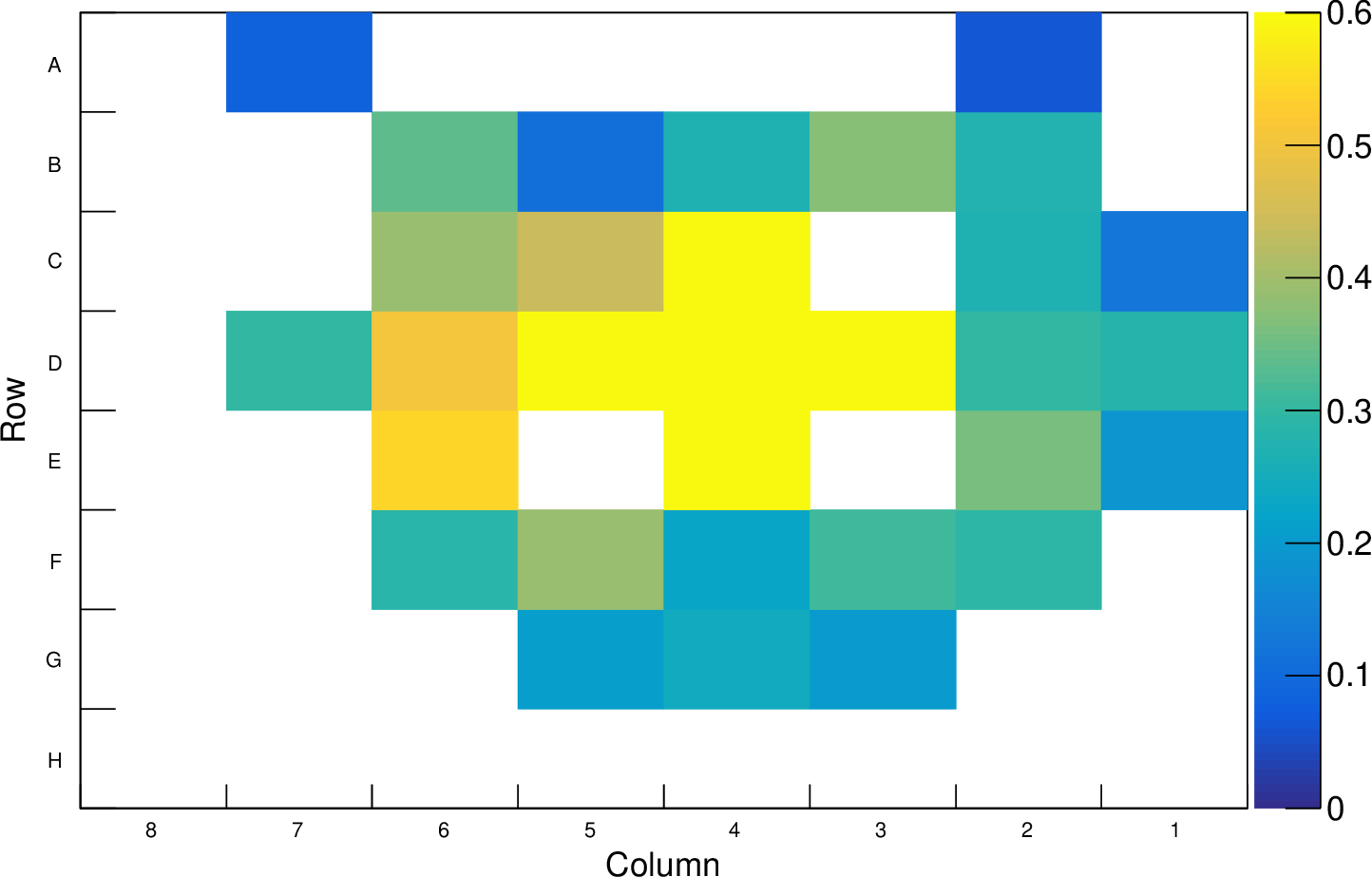}
\caption{\label{fig:qdc_map}Averaged charge (in pC) collected by the readout pads for runs without (left) and with (right) the acrylic filter. The beam spot is located at pad D4. The PEs from the conversion of the Cherenkov photons generated in the radiator populate a 5$\times$5 pad square around the beam spot and, when the acrylic filter is not used, they marginally populate also the more external pads, as demonstrated in pads G3-5 (Sec.~\ref{sec:mc_sim}).
}
\end{center}
\end{figure}

The Cherenkov rings measured by the LAPPD in our standard configuration and without the acrylic filter are shown in Fig.~\ref{fig:qdc_map} together with the beam spot. The beam spot signal was mainly detected in pad D4. Since the triggering beam monitor had dimensions of 5$\times$5 mm$^2$, the center of the beam spot ring was located approximately at the center of that pad. 
The black tape applied on the LAPPD window at the beam spot pad allowed us to suppress the beam spot signal from the expected $\sim$179~PE (Sec.~\ref{sec:mc_sim}) to slightly more than 4~PE. 
The suppression rate of the beam spot light exhibited some variations during the data taking period, probably because the adhesion of the optical grease was varying with time. The Cherenkov light produced by the beam particle in the 5~mm fused silica LAPPD window produced a backward reflected ring of the radius $\sim$15~mm. Therefore, a small portion of this ring was not detected in the pad D4, but it was detected in the four nearby pads. 
\par
The ring of the Cherenkov light from the lens radiator was located at a radius of $\sim$60~mm, corresponding to a distance of 2.4 pads from the center. When the acrylic filter was installed, the average collected charge in these ring pads was lower than 1~SPE, allowing us to conclude that the mean number of PE was $<$1. The mean number of PEs estimated from the data taken with the acrylic filter was $\sim$0.5~PEs. This 
number was four times smaller that the Geant4 expectations of 2 PEs. Nevertheless, the ratio between the mean number of PEs in the runs without and with acrylic filter, a factor of 6, was in a good agreement with the Geant4 simulations. The few readout pads that were located outside the lens Cherenkov ring showed a mean charge three times smaller than the ring pads.

\begin{figure}[!ht]
\begin{center}
\includegraphics[scale=0.27, angle=270]{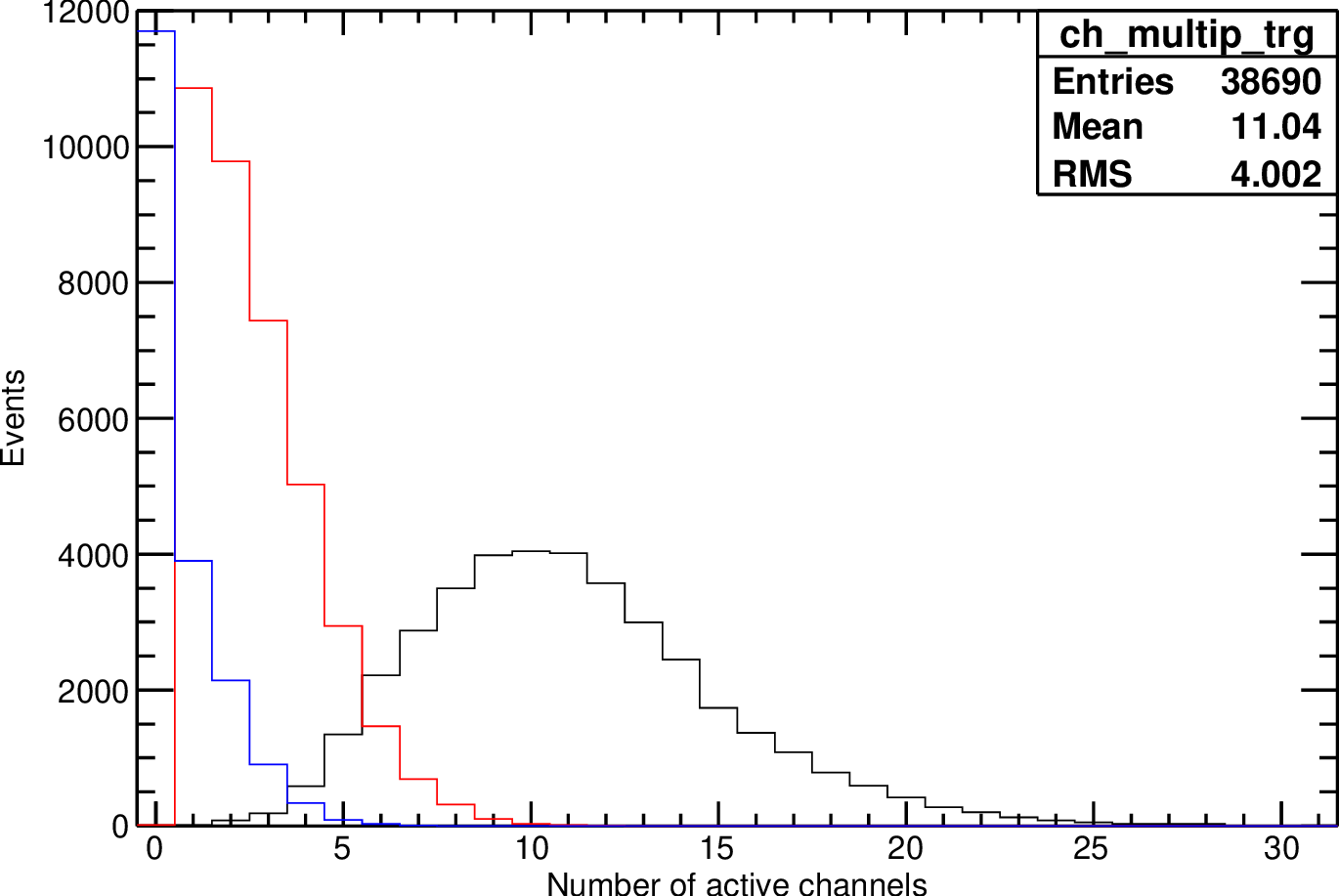}
\caption{\label{fig:hit_multiplicity}Multiplicity of hits above a minimal threshold of 0.1 pC. The black histogram shows all hits above threshold, while the red histogram represents filtered hits, excluding waveforms affected by the cross-talk. The blue histogram represents hit multiplicity in the LED calibration run.}
\end{center}
\end{figure}

In each event the number of hits seen by the LAPPD was on average around 11, as shown in Fig.~\ref{fig:hit_multiplicity}. This provides a realistic simulation of a narrow Cherenkov ring signal in a gas radiator based RICH detector. However, many hits were distorted by the cross talk and only about 24\% of them were selected for the timing analysis.
The multiplicity of selected hits was around 2.7, as shown by the red histogram.
Instead, in the LED calibration run the average multiplicity was around 0.7,
providing essentially pad-by-pad independent measurements.
%
\section{LAPPD Cross-Talk}\label{sec:x_talk}

\begin{figure}[!ht]
\begin{center}
\includegraphics[scale=0.25, angle=270]{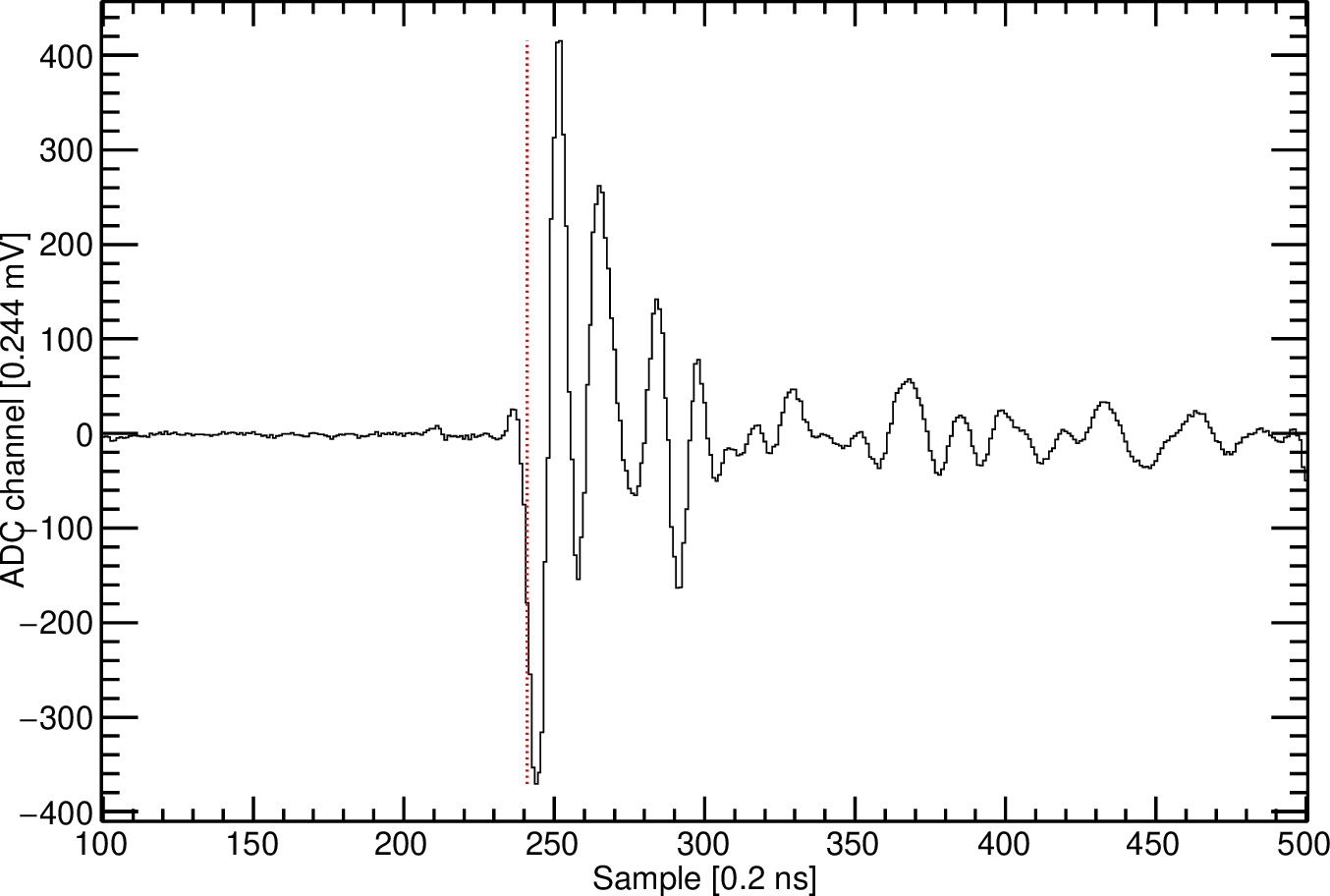}~%
\includegraphics[scale=0.25, angle=270]{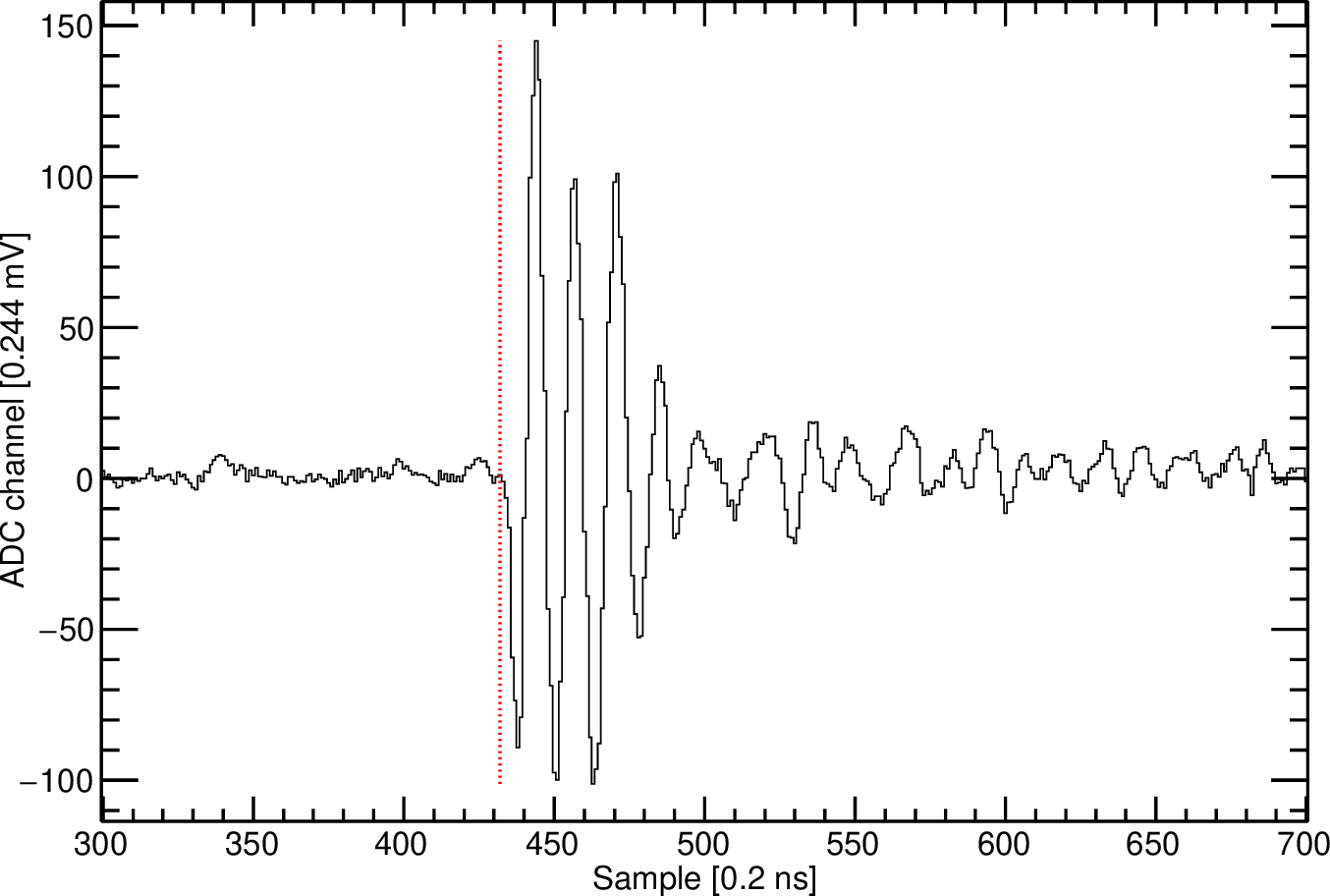}
\caption{\label{fig:wf_xtalk}Waveforms of cross-talk signals acquired during the beamtest obscuring the LAPPD face with a black sheet (left) and in laboratory using a focused pulsed laser source.
Red dotted lines indicate the expected signal start sample.
}
\end{center}
\end{figure}

Large signals in any LAPPD pad led to cross-talk signals in all the other pads. This effect was investigated in dedicated data taking runs by installing an absorptive black sheet between the lens and the LAPPD window. In this configuration, the only Cherenkov signals were those generated by the beam spot, while all the pad off the beam center were populated only by the cross-talk signals.
These signals had approximately a dumped oscillator shape (Fig.~\ref{fig:wf_xtalk}), with the first peak of the amplitude opposite in sign respect to the normal PE signals. The amplitude of these cross-talk signals was approximately constant over the entire LAPPD readout plane.
We reproduced these conditions also in the laboratory with the pulsed laser source by using intense laser pulses illuminating one single pad, while measuring simultaneously the signals in other pads, as shown in Fig.~\ref{fig:wf_xtalk}.
This cross-talk feature represents a potential problem for the usage of AC-coupled LAPPDs, when both large signals by thorough-going ionizing particles and small signals by SPEs are present.

%
%


%
%


%
%

\section{LAPPD Timing Characterization}\label{sec:det_time}


%
%

The LAPPD timing was measured as a difference between the LAPPD signal arrival time
and the arrival time of the Hamamatsu MCP-PMT signal produced by the hadron crossing its window. In this estimate we neglected the timing uncertainty of the Hamamatsu MCP-PMT, which was shown to be $\sim$6~ps~\cite{hamamatsu_MCP}, namely much smaller than the LAPPD resolution we aim to measure.

\subsection{Signal selection}\label{sec:signal-selection}
Waveforms with signals above 25~mV arriving within $\pm$2~ns from SciFi beam monitor signal
were selected for the analysis. In order to suppress the signals contaminated by the cross-talk, the wave shape before and after the main peak was checked to reject the waveforms with dumped oscillator shape.

\subsection{Signal timing}\label{sec:signal-timing}
\begin{figure}[!ht]
\begin{center}
\includegraphics[scale=0.27, angle=270]{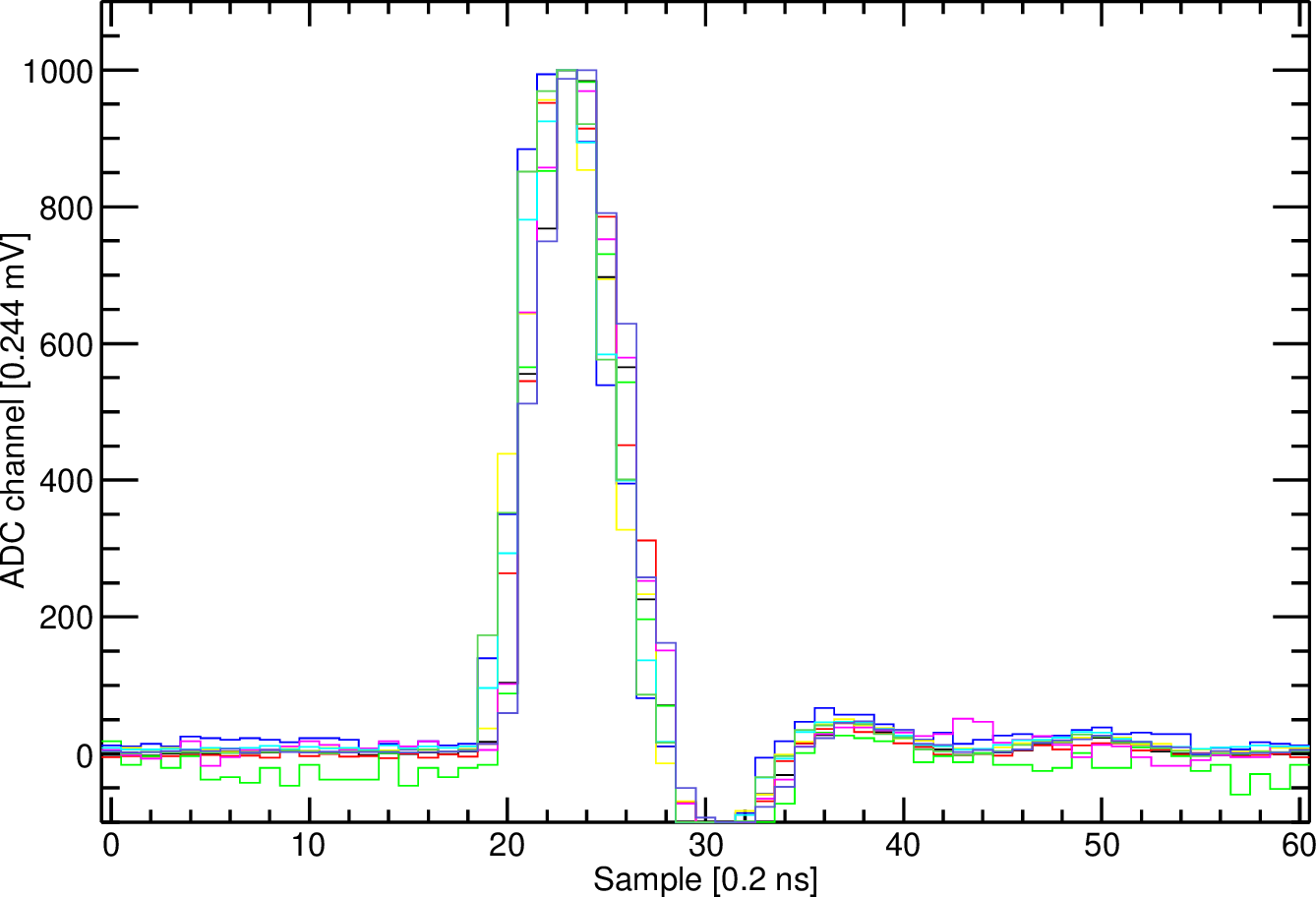}~%
\includegraphics[scale=0.27, angle=270]{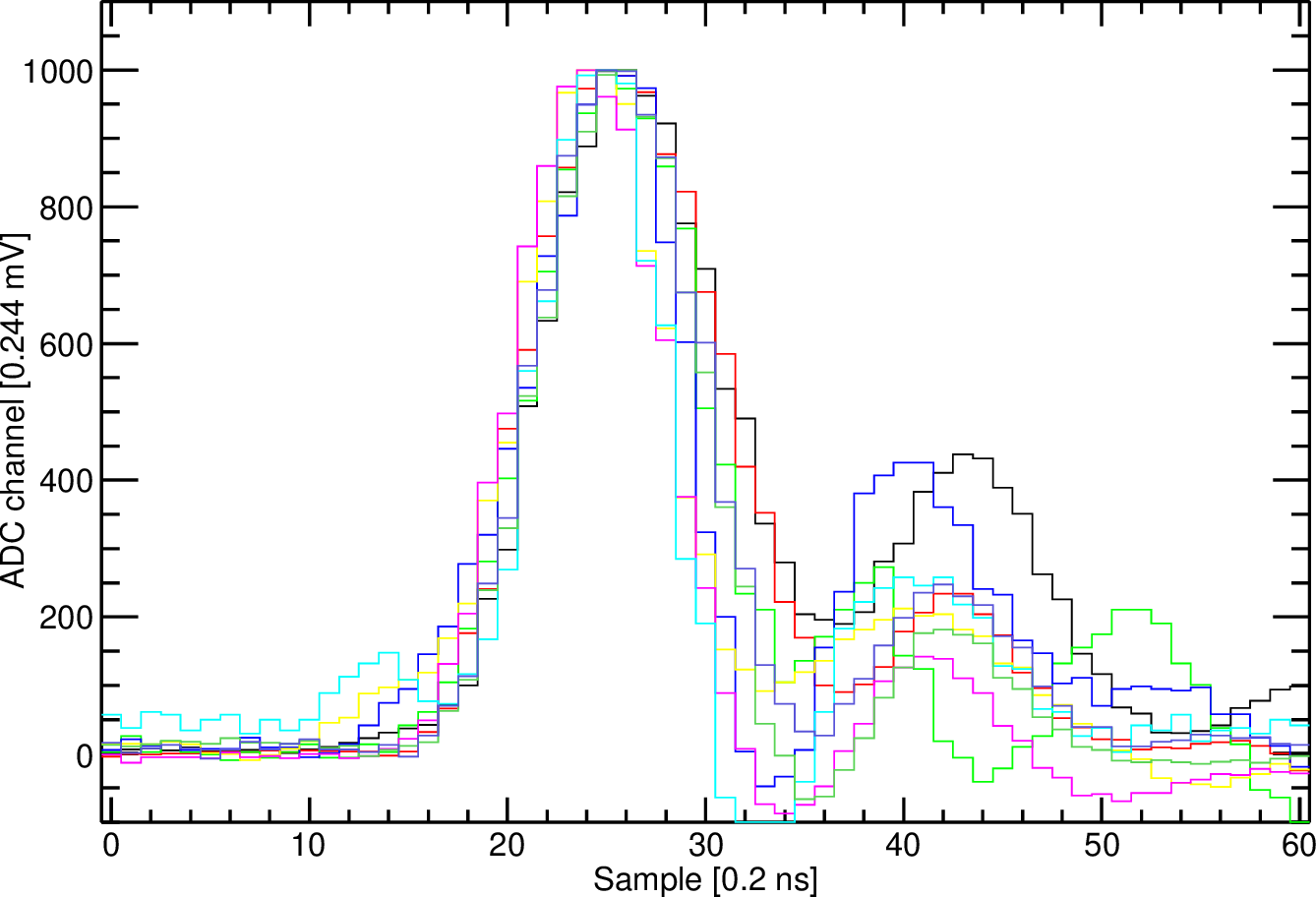}
\caption{\label{fig:pulse_shape}Shapes of the digitized signals from the Hamamatsu MCP-PMT (left) and from a LAPPD channel (right).}
\end{center}
\end{figure}

The time was obtained from the digitized waveforms of the signals, examples of which are shown in Fig.~\ref{fig:pulse_shape}. The rising edge of the signal was fit by a linear function in the range of  (10~-~90)\% of the peak height.
The point of the fit-line at 50\% peak height was taken as the signal time. 
For the Hamamatsu MCP-PMT signals, the rise time was found to be about 0.4~ns. This value was not due to the intrinsic MCP-PMT rise time (expected to be 0.16~ns), but it was limited by 0.5~GHz 
analog bandwidth
of the V1742 digitizer (expected to be 0.44~ns). In case of the LAPPD channels the risetime varies in the range from 0.67 to 0.85~ns. These values were clearly larger than the digitizer average bandwidth limitations and three times larger than one would expect from the pad capacitance.
\par
Timing measurements obtained as explained above, show an amplitude dependence both for the Hamamatsu MCP-PMT and the LAPPD signals, probably due to a non-linear peak rising edge.  This effect has been measured and corrected for. The dependence of the mean time difference of the Hamamatsu MCP-PMT signal amplitude is approximated  linear with -0.2~ps/mV slope coefficient. In the covered amplitude range (400-800~mV), the correction spans the range 0-80~ps. Similarly,  the dependence of the mean time difference of LAPPD signal amplitude is approximated as linear with 0.1~ps/mV slope coefficient and the corresponding range of the correction is 0-60~ps.
Applying amplitude dependent corrections, the timing resolution (Sec.~\ref{sec:timing-distribution-TTS}) improves on average by 6\%, namely subtracting in quadrature 20-60~ps depending on the signal pulse height.
%

\subsection{Timing distributions and the measured transit-time spread}\label{sec:timing-distribution-TTS}
\begin{figure}[!ht]
\begin{center}
\includegraphics[scale=0.27, angle=270]{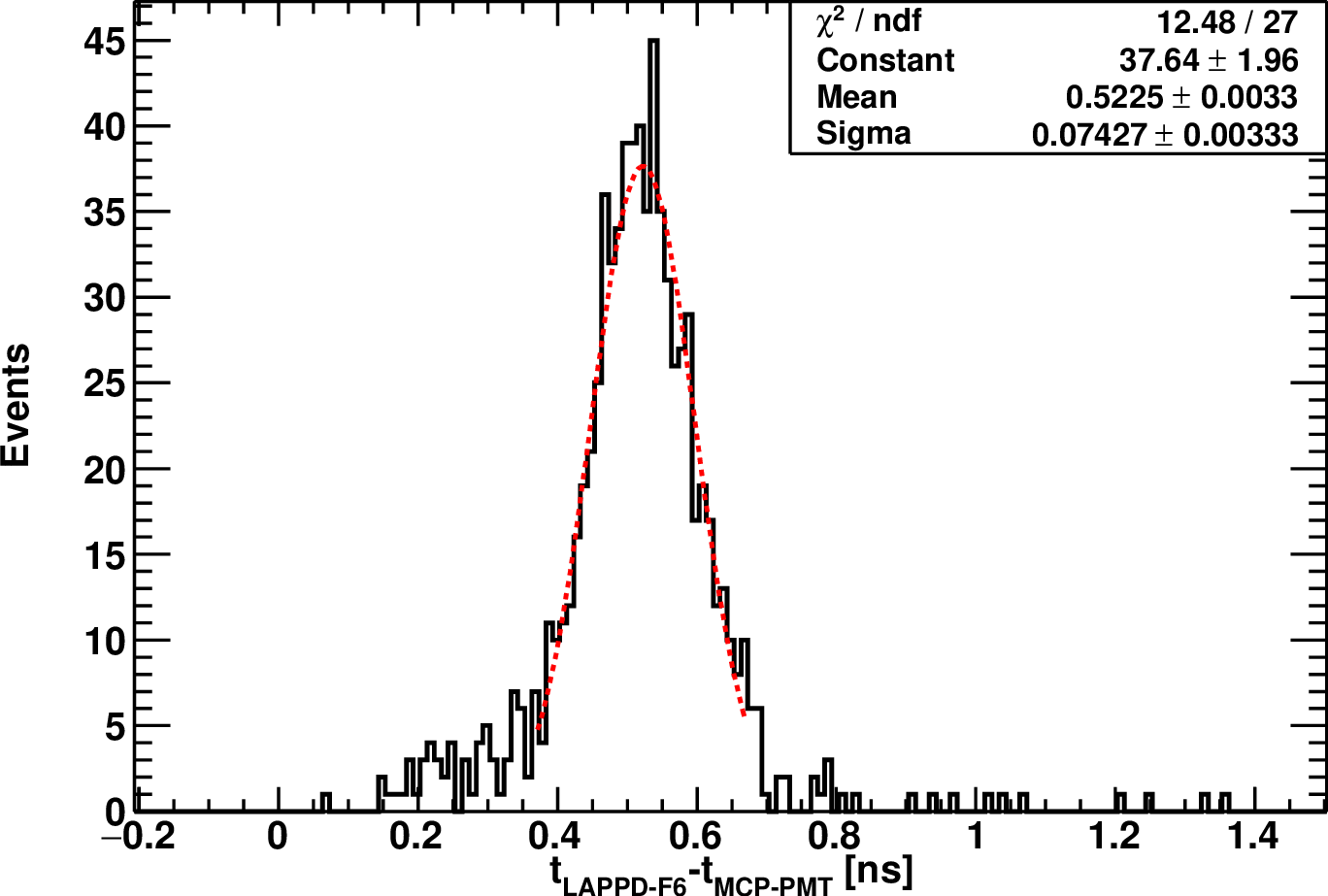}~%
\includegraphics[scale=0.27, angle=270]{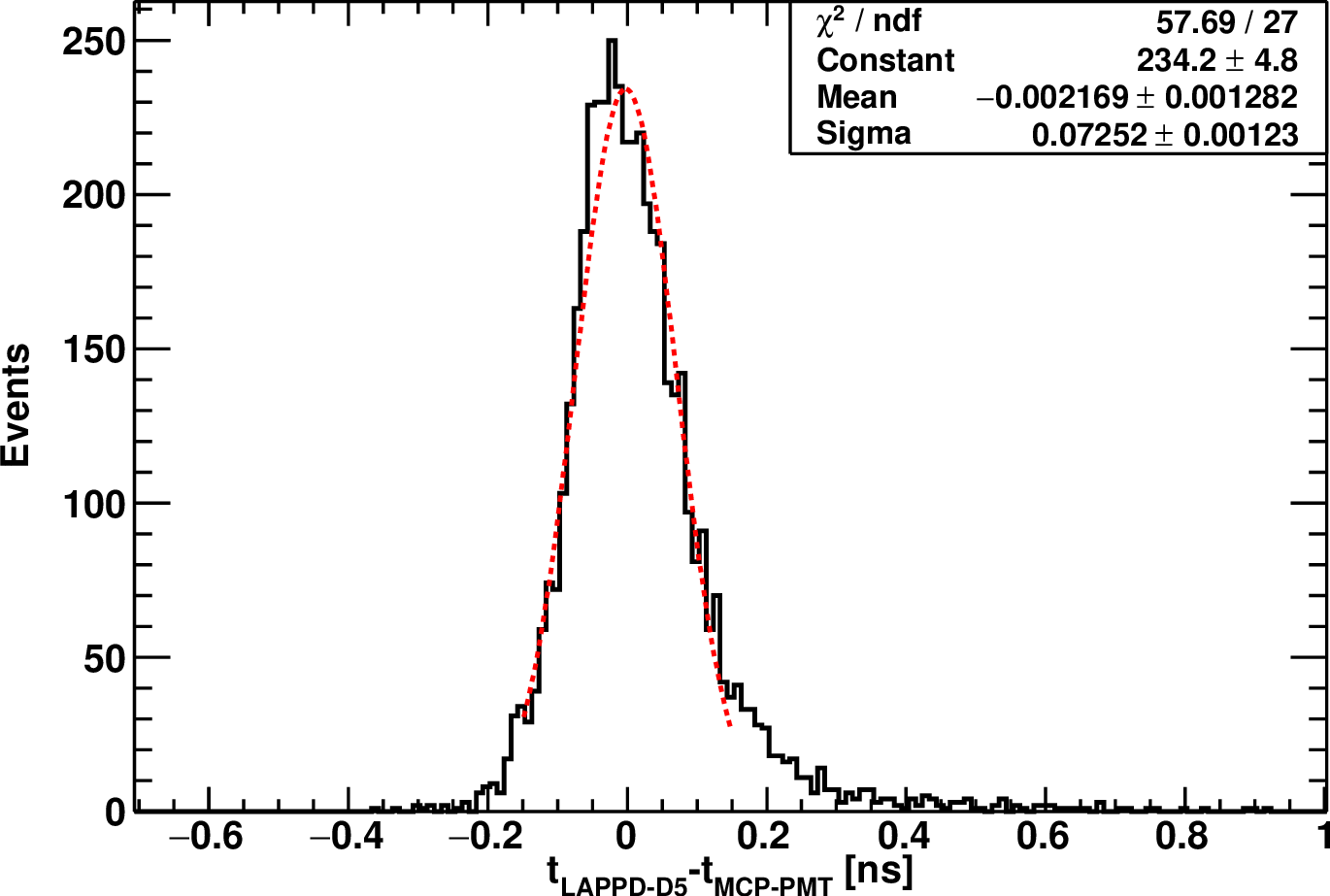}
\caption{\label{fig:dt_lappd_mcp_all}Time difference between LAPPD pad F6 (left) and pad D5 (right) and Hamamatsu MCP-PMT. The pad D5 is close to the beam spot. Gaussian fit is shown with its results.}
\end{center}
\end{figure}

The time difference distributions in two different LAPPD pads are shown in Fig.~\ref{fig:dt_lappd_mcp_all}. The difference in statistics is due to the proximity of pad D5 to the beam spot. 
Also the signal in pad D5 was 
expected to be displaced
with respect to other
channels by 0.52~ns because this was a fraction of the direct beam spot signal shared with the nearby pads. Differently, the Cherenkov photons from the radiator lens reaching pad F6 have an extra path length. 
Geant4 simulations showed that the time delay between Cherenkov photons due to the beam impact on LAPPD window and in the aspheric lens is about 0.5~ns, in a good agreement with our observation.


%
%

The resolution of the time difference distribution is our estimator of the Transit-Time Spread (TTS). 
These distributions have approximately Gaussian shape in the region of the peak. Therefore, we estimated the timing resolution by a Gaussian fit of the peak within $\pm$2$\sigma$.
The obtained $\sigma$-values vary from 75~ps to 120~ps with a fit uncertainty $<$10~ps;
the weighted average over all pads is 87~ps.

The time difference distributions between different LAPPD pads exhibit a resolution approximately enhanced by the factor of $\sqrt{2}$, confirming that the contribution of the Hamamatsu MCP-PMT timing resolution to the LAPPD measured resolution is negligible.


%
%



\subsection{Amplitude dependence of the transit time spread}\label{sec:amplitude-dependence}
\begin{figure}[!ht]
\begin{center}
\includegraphics[scale=0.25, angle=270]{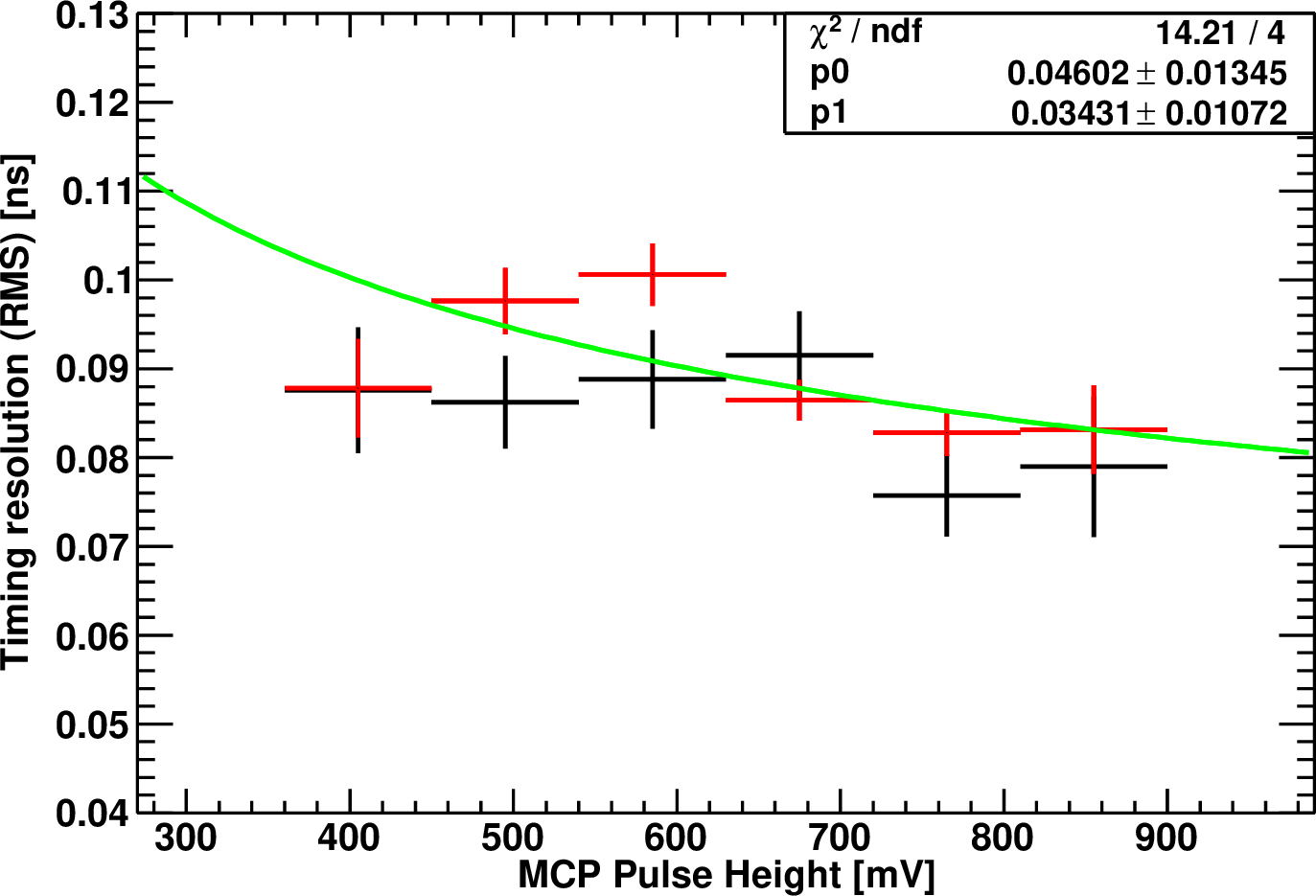}~%
\includegraphics[scale=0.25, angle=270]{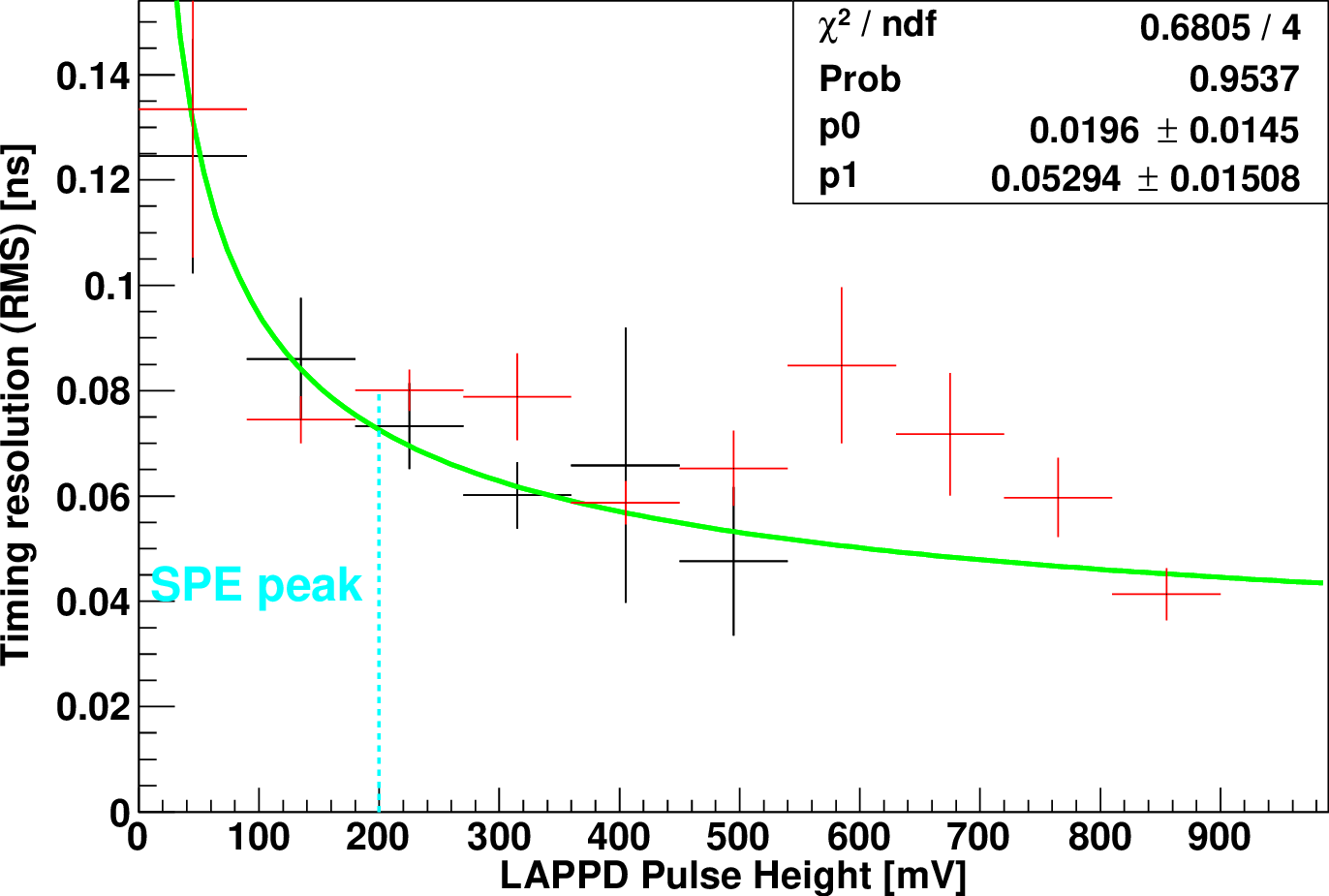}
\caption{\label{fig:cmp_rms_dt}Dependence of the timing resolution (RMS) on the Hamamatsu MCP-PMT (left) and the LAPPD (right) pulse height. Data from pad F6 detecting single photons from the lens radiator (black) and from pad D5 measuring the beam signal (red) are overlapped for comparison. The green lines show a fit to the data with Eq.~\ref{eq:rms_v_fit}. Only the data from pad F6 are fitted in the LAPPD case (right).}
\end{center}
\end{figure}

%
%
%
%
For the Hamamatsu MCP-PMT,  no significant timing distribution dependency from the signal amplitude was observed (Fig.~\ref{fig:cmp_rms_dt}, left), confirming that Hamamatsu MCP-PMT timing resolution contribution was negligible. 
\par
A relevant dependence of the timing resolution on the LAPPD signal amplitude is observed  (Fig.~\ref{fig:cmp_rms_dt}, right), similar for the single PE from the lens radiator and in the beam spot pads. 
\par
In both cases the resolution was fitted using the following functional form:

%
%
\begin{equation}\label{eq:rms_v_fit}
\sigma_t = p_0 + \frac{p_1}{\sqrt{V_{peak}/1 V}}  ~.
\end{equation}
For the Hamamatsu MCP-PMT amplitude dependence, the $p_1$ coefficient is compatible with zero within three standard deviations. For the LAPPD amplitude dependence, the second term in Eq.~\ref{eq:rms_v_fit} is very relevant. The constant resolution term from the fit  is $\sim$20~ps, in a good agreement with the estimated 18~ps systematic uncertainty (Sec.~\ref{sec:systematics}).

%
%

%
%
%

\subsection{Dependence of the timing response from the LAPPD operating voltages}\label{operating-voltages}
\begin{figure}[!ht]
\begin{center}
\includegraphics[scale=0.27, angle=270]{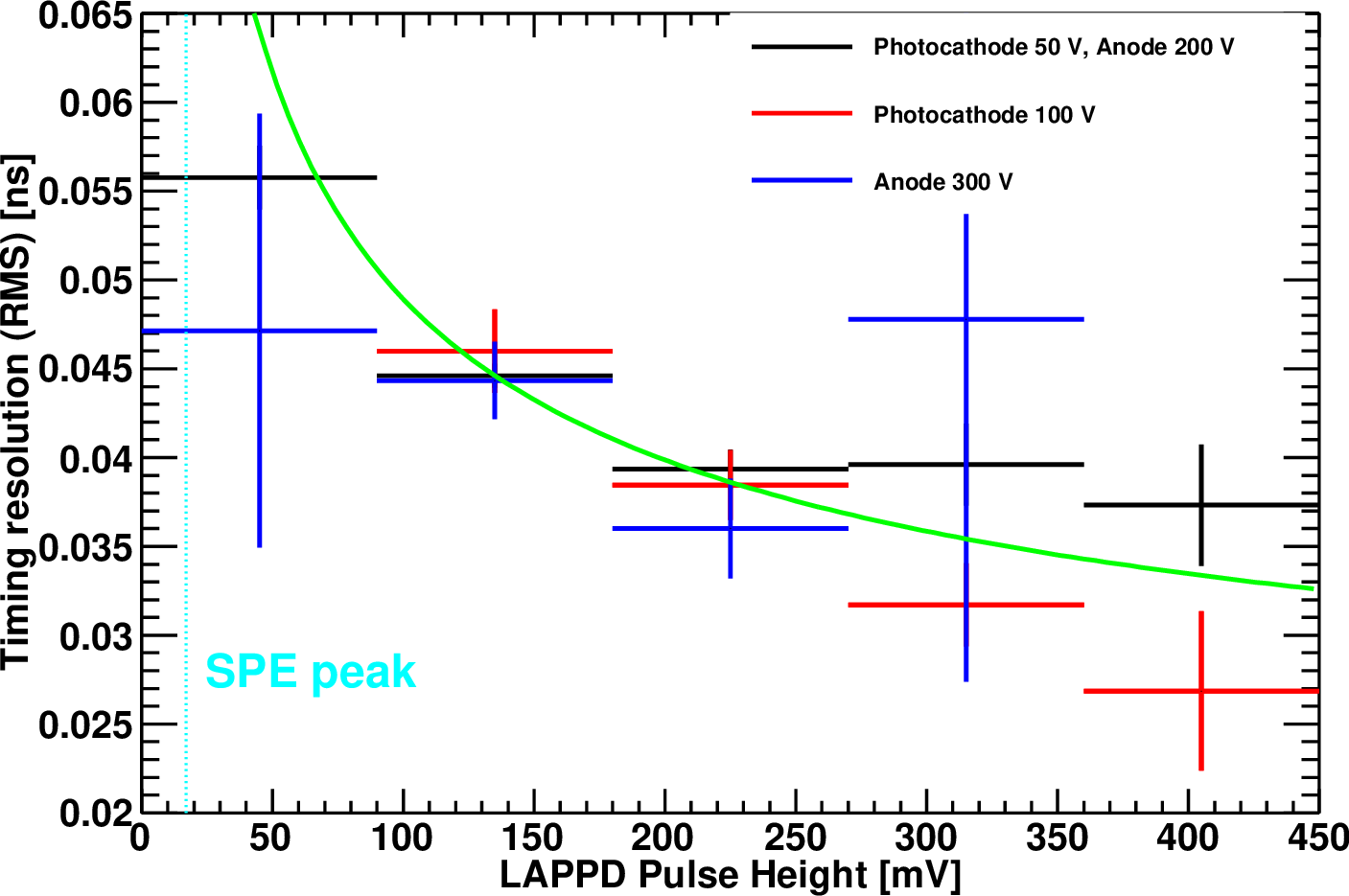}
\caption{\label{fig:cmp_rms_dt_pc_anode}Dependence of timing resolution (RMS) on the LAPPD pulse height in pad D5 at voltages: Photocathode 50 V and Anode 200 V (black), with increased Photocathode voltage to 100 V (red) and with increased Anode voltage to 300 V (blue). A green solid line shows a parameterization from Eq.~\ref{eq:rms_v_fit}
using calculated overall timing uncertainty $p_0=18$~ps and SPE resolution $p_1=75$~ps.
}
\end{center}
\end{figure}

The timing resolution dependence on the LAPPD operating point have been addressed collecting three data samples varying the photocathode-to-first-MCP voltage  and the second-MCP-to-anode voltage. 
The measured timing resolutions for the beam spot pad D5 are shown in Fig.~\ref{fig:cmp_rms_dt_pc_anode}. No significant improvement is obtained increasing both voltages, because the contribution that can come from these voltage settings is negligible already at the lowest voltage setting we used.  Although, the increased photocathode-to-first-MCP voltage data show a slightly better resolution for high amplitude signals, reaching (27$\pm$4)~ps at 400~mV. This value gives the best resolution achieved during the present beam test. This study was performed by absorption only a modest fraction of the beam spot light in order to increase the number of PEs contributing to the beam signal and at lower bias voltages of the LAPPD MCPs: 750 and 800~V, respectively. At these bias voltages, the single PE amplitude was measured to be $\sim$17~mV. Using this calibration, the best resolution point corresponds to about 23~PEs. 
Adding 80~ps single PE resolution, scaled with the square root of the number of PEs, and the estimated 18~ps setup uncertainty (Sec.~\ref{sec:systematics}) in quadrature, we obtain 25~ps estimate for this signal amplitude, in a good agreement with the measured value. The agreement confirms the reliability of our system in measuring timing resolutions down to at least $\sim$25~ps.
\par
In the laboratory, using a pulsed laser source with 50~ps FWHM, we reached 44~ps timing resolution, corresponding to 39~ps RMS.
\par
The same data indicates that the number of generated PEs has a major effect on the timing resolution, as expected from statistical considerations.
In fact, the first data point in Fig.~\ref{fig:cmp_rms_dt_pc_anode}, corresponding to approximately 2.5~PE at a ten times lower gain, can be compared to the point at 500~mV in Fig.\ref{fig:cmp_rms_dt}, left. Both measurements show the timing resolution of about 60~ps with 10\% uncertainties. In our measurements, the effect of operating MCPs at a lower bias voltage, therefore decreasing the average gain, is negligible with respect to other uncertainties. 


%
%

\subsection{Systematic uncertainty}\label{sec:systematics}
The major source of systematic uncertainty in the timing resolution potentially comes from the signal distortion due to the cross-talk generated in the pads with a direct hadron impact, because of the substantially large signals produced there. This contribution appears mostly for the low amplitude signals (below  single PE peak) and can be roughly estimated from the comparison between timing resolutions in different pads and different data taking conditions. While in the low multiplicity and low amplitude calibration runs all pads exhibit the same signal rise time, in the beam test some pads exhibited significantly (+20\% equivalent to +2.5 RMS) larger rise time. Correspondingly, in the beam test, these pads showed a timing resolution increased by 40\%, namely $\sim$40~ps with respect to the average of the other pads. This uncertainty is pad-dependent; therefore, we didn't include it into the final results, for which we selected only the pads not affected by the rise time increase. We assume that the residual systematic uncertainty from this source is negligible.
\par

\begin{figure}[!ht]
\begin{center}
\includegraphics[scale=0.27, angle=270]{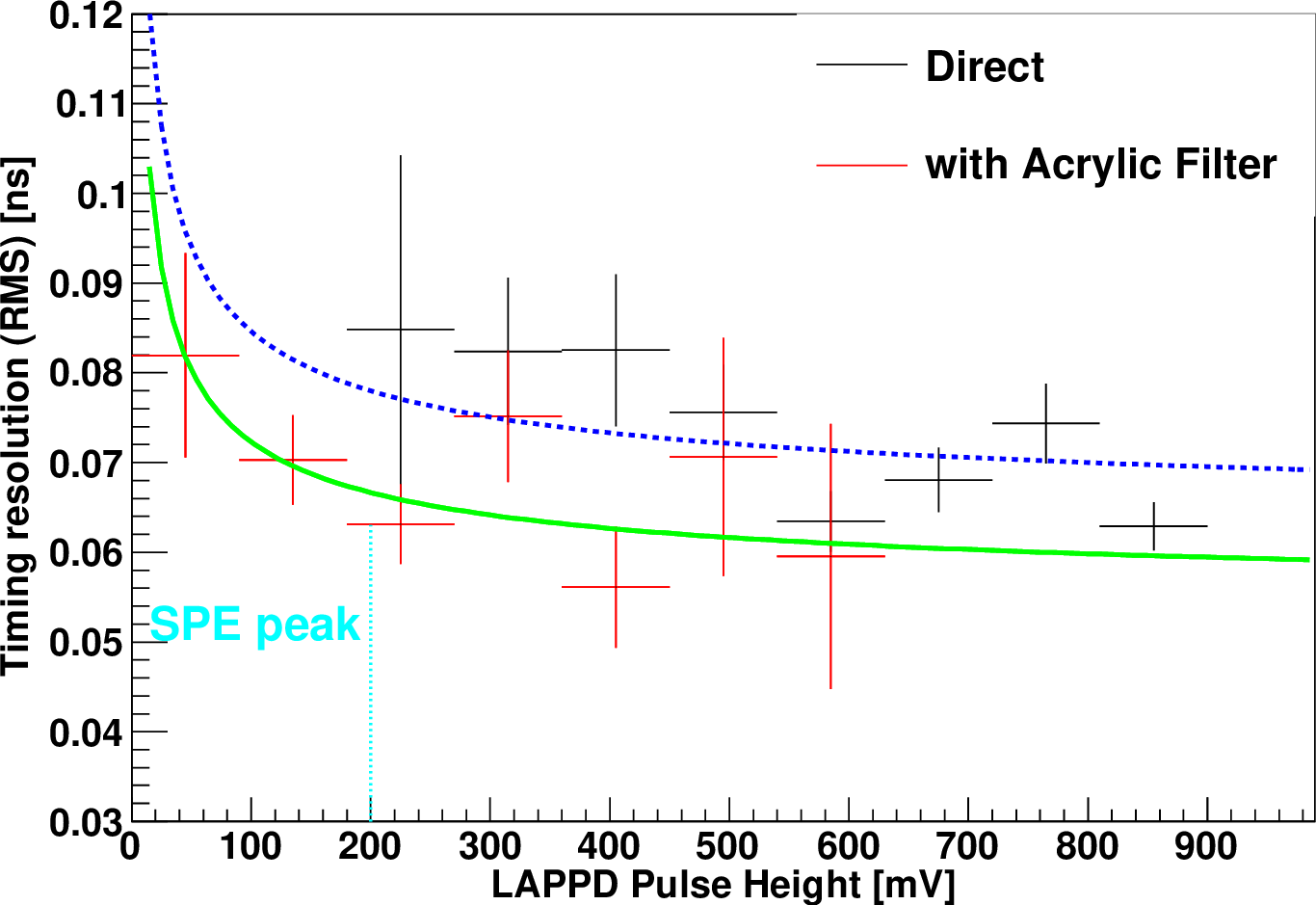}
\caption{\label{fig:cmp_rms_dt_af}Comparison of the timing resolutions (RMS) measured with (red histogram) and without (black histogram) acrylic filter in pad F6.
}
\end{center}
\end{figure}

The chromatic dispersion contribution was studied by comparing the resolutions obtained with and without the acrylic filter in front of the LAPPD. The quartz lens is transparent at least down to 200~nm wavelength, matching the LAPPD fused silica window transmission spectrum and the Ka$_2$NaSb photocathode sensitivity. Installation of the acrylic filter allowed us to limit the wavelength to $>$400~nm, reducing also the average number of photo-electrons per pad. The acrylic filter improves the timing resolution of about 10~ps as shown in Fig.~\ref{fig:cmp_rms_dt_af}. This finding is in agreement with the Geant4 simulations. While the quartz refractive index variation in the range 200-400~nm is about 4.6\% and it is just 0.7\% between 400~nm and 550~nm,  we can estimate the remaining timing resolution uncertainty due to the chromatic dispersion to be about 1.5~ps, included in the overall setup uncertainties (as explained below).

\par
The mean Signal-to-Noise ratios (S/N) in LAPPD and Hamamatsu MCP-PMT were about 130 and 400, respectively. Therefore, including also the known TTS of the Hamamatsu MCP-PMT, these two terms add 5~ps and 10~ps respectively to the overall resolution.
\par
Monte Carlo simulations performed in Geant4 allowed us to estimate the overall contribution of the setup geometry and the chromatic dispersion. Assuming the beam spot extent as defined by the beam counter (5$\times$5~mm$^2$) and the use of the acrylic filter, we obtained a contribution to the timing RMS of 8.3~ps.
\par
The readout pad size of 1 inch squared is a source of another non-negligible contribution to the signal arriving time with RMS of the order of 12~ps.

Assuming that all the contributions above are independent and combining them in quadrature, we obtain an uncertainty of 18~ps non dependent on the LAPPD signal amplitude and a contribution dependent on the amplitude as in the second term in Eq.~\ref{eq:rms_v_fit} with $p_1$~=~5~ps.

\section{Conclusions}\label{sec:conclusions}
In the present article we described a first measurement of the SPE timing resolution of a Generation II (capacitively coupled) 20~$\mu$m pore LAPPD detecting Cherenkov photons produced in a fused silica lens radiator by the thorough-going hadrons. Cherenkov photons generated by the beam particle in the LAPPD 5~mm thick fused silica window were detected as well.  The measurement was performed at the CERN PS T10 test beam line with hadrons of different momenta in the range 4~-~10~GeV/c.
\par
The measured LAPPD timing resolution varied from pad to pad, also related to limitations in the synchronization of the electronic readout channels. On average for the pads read out with the best synchronized electronics channels, the resolution is $\sim$80~ps. This value is much larger than all experimental uncertainties related to the light arrival time at the LAPPD window and to the electrical signal readout and reconstruction, whose combined effect is estimated to be $\sim$18~ps. 
The best achieved resolution for multiple PE signals is 27~ps, close to the expected figure of 25~ps, demonstrating the capability of our approach in measuring time resolutions as fine as 25-30~ps.
Therefore, we could conclude that the obtained timing resolution should be attributed to the intrinsic TTS of the Generation II LAPPD with 20~$\mu$m pores, when operated at the biasing voltages we used.
\par
Significant cross-talk affecting all the readout pads was observed, both at the test beam and reproduced in the laboratory with a pulsed laser source.
\par
Reduced TTS is expected in the next generation LAPPDs and in HRPPDs\footnote{HRPPDs are sensors by INCOM with similar architecture as LAPPDs, reduced active surface of 100$\times$100~mm$^2$ and requiring no support for the fused silica window} with 10~$\mu$m pore and shorter gaps. The DC-coupling readout might help in reducing the cross-talk effect.

\section*{Acknowledgements}
The authors would like to acknowledge the excellent support provided during the experiment by the staff and technical services of CERN PS T10 beamline and the kind hospitality of the ePIC dRICH group, main user of the test beam used for our studies.
Our special gratitude goes to Eraldo Oliveri for the usage of the Hamamatsu MCP-PMT, to Berthold Jenninger for providing the CAEN DT1415 power supply and to Vincenzo Vagnoni for assistance with digitizer timing calibrations.
Authors also thank Incom for their assistance from remote, and Emmanuel Rauly for his advises on the HPS amplifier.
\par
These studies have been partially supported by the eRD110 project of the EIC R\&D program and by the the European Union’s Horizon 2020 Research and Innovation programme under the Grant Agreement AIDAinnova - No 101004761.
\bibliographystyle{elsarticle-num}
\bibliography{lappd_t}

\begin{thebibliography}{10}
\expandafter\ifx\csname url\endcsname\relax
  \def\url#1{\texttt{#1}}\fi
\expandafter\ifx\csname urlprefix\endcsname\relax\def\urlprefix{URL }\fi
\expandafter\ifx\csname href\endcsname\relax
  \def\href#1#2{#2} \def\path#1{#1}\fi

\bibitem{Farnsworth}
P.~Farnsworth, {Electron Multiplier}, {US Patent 1,969,399} (1930).

\bibitem{Goodrich}
G.~Goodrich, W.~Wiley, Continuous channel electron multiplier, Rev. Sci.
  Instrum. 33 (1962) 761.

\bibitem{Wiza}
J.~Wiza, Microchannel plate detectors, Nucl. Instr. and Meth. 162 (1979) 587.

\bibitem{Beaulieu_1}
D.~Beaulieu, et~al., Nano-engineered ultra high gain microchannel plates, Nucl.
  Instr. and Meth. A 607 (2009) 81.

\bibitem{Beaulieu_2}
D.~Beaulieu, et~al., Plastic microchannel plates with nano-engineered films,
  Nucl. Instr. and Meth. A 633 (2011) S59.

\bibitem{lappd_intro}
A.~Lyashenko, et~al., Performance of large area picosecond photo-detectors,
  Nucl. Instr. and Meth. A 958 (2020) 162834.

\bibitem{advance_lappd}
C.~A. Craven, et~al., {Recent Advances in Large Area Micro-channel Plates and
  LAPPD}, Springer Proc.Phys. 213 (2018) 319.

\bibitem{shin_lappd}
S.~Shin, et~al., \href{https://arxiv.org/abs/2212.03208}{{Advances in the Large
  Area Picosecond Photo-Detector (LAPPD): 8x8 MCP-PMT with Capacitively Coupled
  Readout}}, 2022.
\newline\urlprefix\url{https://arxiv.org/abs/2212.03208}

\bibitem{annie}
\href{https://annie.fnal.gov/}{{Annie experiment}}.
\newline\urlprefix\url{https://annie.fnal.gov/}

\bibitem{lappd_time_p}
M.~J. Minot, et~al., Large area picosecond photodetector offers fast timing for
  nuclear physics and medical imaging, Il Nuovo Cim. C 43 (2020) 11.

\bibitem{lubliana_lappd}
A.~Seljak, et~al., {LAPPD operation using ToFPETv2 PETSYS ASIC}, JINST 18
  (2023) C02007, {Contribution to TWEPP-22}.

\bibitem{cherenkov_lappd}
T.~Kaptanoglu, et~al., {Cherenkov and scintillation separation in water-based
  liquid scintillator using an LAPPD}, Eur.Phys.J. C 82 (2022) 169.

\bibitem{LAPPD_Workshop_1}
\href{https://indico.bnl.gov/event/15059/}{{1st LAPPD Workshop}} (2022).
\newline\urlprefix\url{https://indico.bnl.gov/event/15059/}

\bibitem{LAPPD_Workshop_2}
\href{https://indico.bnl.gov/event/17475/}{{2nd LAPPD Workshop}} (2022).
\newline\urlprefix\url{https://indico.bnl.gov/event/17475/}

\bibitem{LAPPD_Workshop_3}
\href{https://indico.bnl.gov/event/18642/}{{3nd LAPPD Workshop}} (2023).
\newline\urlprefix\url{https://indico.bnl.gov/event/18642/}

\bibitem{white_paper}
A.~Accardi, et~al., {Electron Ion Collider: The Next QCD Frontier -
  Understanding the glue that binds us all}, Eur. Phys. J. A 52 (2016) 268.

\bibitem{yellow_report}
R.~A. Khalek, et~al., {Science Requirements and Detector Concepts for the
  Electron-Ion Collider: EIC Yellow Report}, Nucl. Instr. and Meth. A 1026
  (2022) 122447.

\bibitem{geant4}
S.~Agostinelli, et~al., Geant4—a simulation toolkit, Nucl. Instr. and Meth.
  506 (2003) 250.

\bibitem{hps_amplifier}
E.~Rauly, G.~Charles,
  \href{https://misportal.jlab.org/mis/physics/hps\_notes/viewFile.cfm/2016-00%
1.pdf?DocumentId=17}{Current sensitive preamplifier used for hps calorimeter},
  hPS-Note 2016-001 (2016).
\newline\urlprefix\url{https://misportal.jlab.org/mis/physics/hps\_notes/viewF%
ile.cfm/2016-001.pdf?DocumentId=17}

\bibitem{v1742}
\href{https://www.caen.it/products/v1742/}{{CAEN SpA, Digitiser v1742.
  Available online:}} (2022).
\newline\urlprefix\url{https://www.caen.it/products/v1742/}

\bibitem{DRS}
R.~H. Ch.~Br\"onnimann, R.~Schnyder, The domino sampling chip: a 1.2 ghz
  waveform sampling cmos chip, Nucl. Instr. and Meth. A 420.

\bibitem{DRS4}
S.~Ritt, et~al., {Application of the DRS Chip for Fast Waveform Digitizing},
  NIMA 623 (2010) 486--488.

\bibitem{DRS4_old}
S.~Ritt, The drs chip: cheap waveform digitizing in the ghz range, Nucl. Instr.
  and Meth. A 518.

\bibitem{ritt_calibration}
S.~R. D.~Stricker-Shaver, et~al., {Novel Calibration Method for Switched
  Capacitor Arrays Enables Time Measurements With Sub-Picosecond Resolution}, A
  IEEE Transactions on Nuclear Science 61 (2014) 3067.

\bibitem{kimetal}
H.~Kim, et~al., A new time calibration method for
  switched-capacitor-array-based waveform samplers, Nucl. Instr. and Meth. A
  767.

\bibitem{drs4_time_calibration}
D.~Stricker-Shaver, S.~Ritt, B.~Pichler, Large area picosecond photodetector
  offers fast timing for nuclear physics and medical imaging, IEEE Transactions
  on Nuclear Science 61 (2014) 3607.

\bibitem{hamamatsu_MCP}
J.~Bortfeldt, et~al., Timing performance of a micro-channel-plate
  photomultiplier tube, Nucl. Instr. and Meth. A 960 (2020) 163592.

\end{thebibliography}

\end{document}